\documentclass[a4paper,floatfix,rmp,twocolumn,showkeys,superscriptaddress]{revtex4}
  \usepackage[latin1]{inputenc}
  \usepackage{graphicx}
  \usepackage{mathtools}
  \setcitestyle{numbers,square,comma}

  \usepackage{amsmath}

\usepackage{multirow}%
\usepackage{amsmath,amssymb,amsfonts}%
\usepackage{amsthm}%
\usepackage{mathrsfs}%
\usepackage{xcolor}%
\usepackage{textcomp}%
\usepackage{booktabs}%

\begin{document}

  \title{Topological communities in complex networks}

  \providecommand{\CNB}{Departamento de Biolog\'ia de Sistemas, Centro Nacional de Biotecnolog\'ia (CSIC), C/ Darwin 3, 28049 Madrid, Spain. }
  \providecommand{\GISC}{Grupo Interdisciplinar de Sistemas Complejos (GISC), Madrid, Spain.}
  
  \author{Lu\'is F Seoane}
    \affiliation{\CNB}
    \affiliation{\GISC}

  \vspace{0.4 cm}
  \begin{abstract}
    \vspace{0.2 cm}

    Most complex systems can be captured by graphs or networks. Networks connect nodes (e.g.\ neurons) through edges (synapses), thus summarizing the system's structure. A popular way of interrogating graphs is community detection, which uncovers sets of geometrically related nodes. {\em Geometric communities} consist of nodes ``closer'' to each other than to others in the graph. Some network features do not depend on node proximity---rather, on them playing similar roles (e.g.\ building bridges) even if located far apart. These features can thus escape proximity-based analyses. We lack a general framework to uncover such features. We introduce {\em topological communities}, an alternative perspective to decomposing graphs. We find clusters that describe a network as much as classical communities, yet are missed by current techniques. In our framework, each graph guides our attention to its relevant features, whether geometric or topological. Our analysis complements existing ones,  and could be a default method to study networks confronted without prior knowledge. Classical community detection has bolstered our understanding of biological, neural, or social systems; yet it is only half the story. Topological communities promise deep insights on a wealth of available data. We illustrate this for the global airport network, human connectomes, and others. 

  \end{abstract}


\maketitle

  Network science has revolutionized our understanding of diverse systems ranging from gene regulation \cite{shen2002network, farkas2003topology, luscombe2004genomic, dobrin2004aggregation, palla2005uncovering}; through neural circuitry \cite{chen2008revealing, bullmore2009complex, van2012high, harriger2012rich, betzel2014changes}, ecology \cite{montoya2006ecological, ings2009ecological, bascompte2003nested}, linguistics \cite{corominas2009ontogeny, goni2011semantic, sole2015ambiguity, seoane2018morphospace, corominas2018chromatic}, or technology \cite{valverde2007topology, valverde2015punctuated}; to human \cite{milgram1967small, travers1977experimental,palla2005uncovering}, political \cite{neal2014backbone, andris2015rise, neal2020sign, hohmann2023quantifying}, or economic interactions \cite{onnela2003dynamics}. Networks abstract away such complex systems into a set of nodes or vertices (e.g.\ genes, neurons, species, people, or companies; etc.)\ connected by edges or links (respectively: promotion or inhibition, synapses, predation or mutualism, friendship, or supply dependencies; etc.). These are pair-wise interaction summaries, which often suffice to capture what matters in each system \cite{bialek2007rediscovering}. Interrogating the resulting graphs is much simpler than running detailed models of each case. 

  A common strategy to study networks is hypothesis driven: We suspect that a feature plays an important role (e.g.\ hierarchy \cite{corominas2013origins}, motifs as building blocks \cite{milo2002network}, or a backbone within the human brain \cite{van2012high}), so we set out to find these elements, quantifying node involvement, and measuring graph properties (e.g.\ communication efficiency and cost to bring that brain backbone to light \cite{van2012high}). This requires some prior insight about the system, which we might acquire after visual inspection of the network. Could the graph guide our attention to its outstanding features in a more automated way? That is achieved, within a specific scope, by community detection algorithms, which uncover relevant subgraphs based on proximity criteria---e.g., by grouping nodes that are more connected to each-other (hence closer in a geometric sense) than to the rest of the network \cite{newman2006finding, peixoto2021descriptive}. Let us call such sets of nodes {\em classic} or {\em Geometric Communities} (GC). Some GC stand out visually as a graph is plotted, and human feedback can be considered. But, in general, good algorithms see through the tangled web of connections finding us clusters difficult to see with the naked eye. These communities decompose the network often uncovering functional modules---e.g.\ functional gene relationships \cite{palla2005uncovering}, brain circuits \cite{chen2008revealing}, or ideological political groups \cite{neal2014backbone, andris2015rise, neal2020sign, hohmann2023quantifying}. 

  These two approaches (hypothesis-driven and community detection studies) underlie most network analyses, and are behind the revolution that network science brought about. Might there be a blind-spot that we have not exploited yet? Classic community detection is restricted to finding contiguous sets of nodes. But relevant network features are often distributed---e.g.\ nodes may play similar roles because they act as bridges between communities, or because they constitute a backbone holding the graph together. These functions can be implemented by vertices that are not necessarily close-by, hence would be missed by classic community detection. We might suspect such functionality, as in \cite{van2012high}; but the combinatorial possibilities are staggering and our capacity is limited. Is there a more general, automated way for a network to direct our attention to its most salient features, whether they stem from geographic proximity or from similarity between node types? 

  \begin{figure*}[] 
    \begin{center} 
      \includegraphics[width=\linewidth]{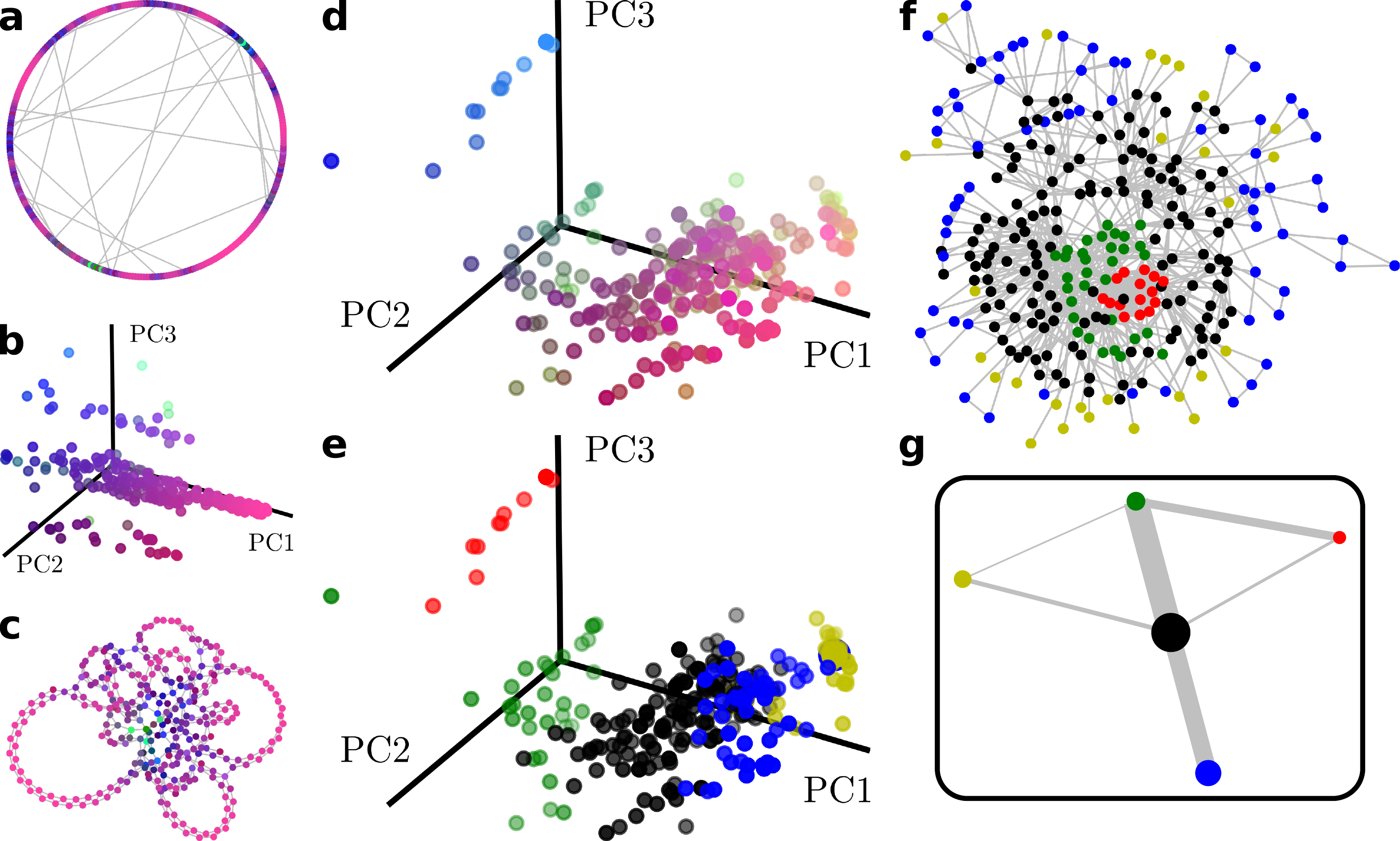}
  
      \caption{{\bf Topological communities in Watts-Strogatz graphs and a collaboration network.} {\bf a} A Watts-Strogatz graph, $\mathcal{G}^{WS}$, show topological defects near its shortcuts. These become apparent by a coding scheme that captures the position of each node in an eigenspace of topological properties ({\bf b}). {\bf c} Same network and color-coded topology in a spring layout. {\bf d} Eigenspace of topological properties of nodes from a collaboration network, $\mathcal{G}^{CNB}$. {\bf e} Distances in this space can be used to define {\em Topological Communities}. Here, five TC are colored distinctly. {\bf f} Projecting back the color-coded communities reveals their roles in the network as a shell (black) that separates a core (green) and a rich club (red) from two distinct peripheries (blue and yellow). {\bf g} These communities induce a network decomposition that can be used to produce a coarse-grained graph that summarizes connections across TC---and highlights, e.g., the role of the shell. }
  
      \label{fig:1}
    \end{center}
  \end{figure*}

  A minimal example is illustrative. In a Watts-Strogatz network \cite{watts1998collective}, $\mathcal{G}^{WS}$, all nodes start as exactly identical, sitting around a circle and each connected to their $k$ nearest neighbors. At this point, whichever topological property we measure on the vertices, they all register the same. Now with a probability $p$ for each edge, we break it and make a shortcut from one of the nodes just separated to another, random one across the graph (Fig.\ \ref{fig:1}{\bf a}). This introduces topological defects: One of the separated neighbors has lost a connection, which is gained in turn by the far-away node. Clustering near the shortcut decreases, since long-range triangles are not completed. Distances across the graph change. Etc. From each node, we measure these and other properties---e.g.\ centralities such as betweenness, closeness, eigenvector; local connectivity such as clustering, $k$-coreness, cliques; cycles associated to each vertex; etc.\ (App.\ \ref{app:1}). We then study how these properties were perturbed by rewiring, and how the earlier topological homogeneity is recovered in nodes further away from shortcuts. Projecting these properties into Principal Components (PC, App.\ \ref{app:2}) reveals how the more homogeneous nodes occupy a limited region (pink tip, Fig.\ \ref{fig:1}{\bf b}), and how the emerging range of topologically distinct vertices spreads over this eigenspace. Using red (PC-$1$), green (PC-$2$), and blue (PC-$3$) to code for position respectively along the first three components (App.\ \ref{app:2}), we map this topological diversity back into the $\mathcal{G}^{WS}$ circle graph (Fig.\ \ref{fig:1}{\bf a}) or another suitable layout (Fig.\ \ref{fig:1}{\bf c}). Nodes far from topological accidents (i.e.\ similar to themselves before rewiring) stand out along the red-coded PC-$1$, and thus they color the network in pink stretches of very regular, almost grid-like structures (Fig.\ \ref{fig:1}{\bf c}). Properties typical of a grid (e.g.\ square clustering, long average distance to other vertices) have bigger loadings on PC-$1$ in this example. Nodes closer to a shortcut score high in betweenness centrality instead. In Fig.\ \ref{fig:1}{\bf a-c}, these later vertices acquire more bluish (PC $3$) and eventually greener (PC $2$) hues, indicating that these PC correlate with betweenness and other defining aspects of shortcuts in Watts-Strogatz graphs.  

  The topological diversity within this network is limited. A more telling example comes from a scientific collaboration graph \cite{manrubia2022report}, $\mathcal{G}^{CNB}$ (App.\ \ref{app:0}). We measured topological properties (again, capturing centrality, local connectivity, cycles, etc.; App.\ \ref{app:1}) of every node in this graph and projected them onto PC eigenspace (Fig.\ \ref{fig:1}{\bf d}). PC in $\mathcal{G}^{CNB}$ are different from those in $\mathcal{G}^{WS}$. In our analysis, each network reveals those features that better explain the topological diversity of its nodes. In this collaboration network, PC-$1$ defines an axis of centrality. More central researchers (in terms of eigenvector, betweenness, closeness, etc.)\ are projected onto smaller values of PC-$1$; while peripheral vertices score higher there (Fig.\ \ref{fig:1}{\bf d} and Sup.\ Fig.\ \ref{fig:SM.3}). This reveals a core-periphery structure that becomes obvious as we use distances in PC-space to define clusters of similar nodes (Fig.\ \ref{fig:1}{\bf e}). We term such clusters {\em Topological Communities} (TC, App.\ \ref{app:3}). TC consist of nodes more similar to each other (in topological terms) than to the rest of the network. Fig.\ \ref{fig:1}{\bf e} shows $5$ TC for $\mathcal{G}^{CNB}$, but TC form a hierarchy that can be explored seeking finer topological detail (Sup.\ Fig.\ \ref{fig:SM.5}). 

  Projecting TC back onto a network layout (Fig.\ \ref{fig:1}{\bf f}) helps clarify their roles. $\mathcal{G}^{CNB}$ consists of a shell (TC-$1$, black) that separates two topologically different sets of peripheral nodes (TC-$4$, blue, and TC-$5$, yellow) from a marked core (TC-$3$, green) alongside a rich club (TC-$2$, red). The rich club consists of scientists very central to the network, who collaborate amply across shell and core, but much more among themselves \cite{manrubia2022report}. Both the core and rich club score high on centrality measurements (low PC-$1$). But they are told apart by PC-$3$, which correlates (among others) with rich-club properties, such us nodes beloging to very large cliques and $k$-cores, and completing a large number of neighbor triangles. One of the peripheries (TC-$5$, yellow), consists of researchers with only one collaborator---they are terminal leaves of the graph, suggesting newcomers to the collaboration network. The other periphery, TC-$4$ (blue), contains researchers with several connections. Both periferies are told apart by PC-$2$ (Fig.\ \ref{fig:1}{\bf e}). Nodes in TC-$4$ are topologically similar to each other (they present akin values of centrality, involvement in cycles, connectivity patterns, etc.). But TC-$4$ is not contiguous---we cannot visit every TC-$4$ node without passing through other TC, prominently the shell (Sup.\ Fig.\ \ref{fig:SM.6}{\bf a}). Hence, despite its topological regularity and marked role within $\mathcal{G}^{CNB}$, TC-$4$ could never be picked up by classic GC. Actually, no $\mathcal{G}^{CNB}$ TC is recovered by GC despite their prominence and the clear network decomposition they entail (Sup.\ Fig.\ \ref{fig:SM.6}{\bf c-d}). 

  Approaches to topological classification exist that divide graphs between core and periphery based on centrality measurements \cite{csermely2013structure, rombach2014core, hebert2016multi}. But a network might not have clear-cut core and periphery---e.g.\ $\mathcal{G}^{WS}$ and others below. A graph's structure might also be more nuanced than captured by a single, monotonously increasing feature---e.g.\ in $\mathcal{G}^{CNB}$ a PC uncovers the core-periphery but other, orthogonal components are needed to extract additional details. The TC framework is both more precise, subtle, and unbiased in identifying cores, peripheries, and further structure---if present. It allows an automated analysis in which each graph guides us to its most salient features, whether their nature depends on centrality or other properties. It acts as a microscope that amplifies each graph's defining structures. Let us turn this approach to more relevant graphs. 

  \begin{figure*}[] 
    \begin{center} 
      \includegraphics[width=\linewidth]{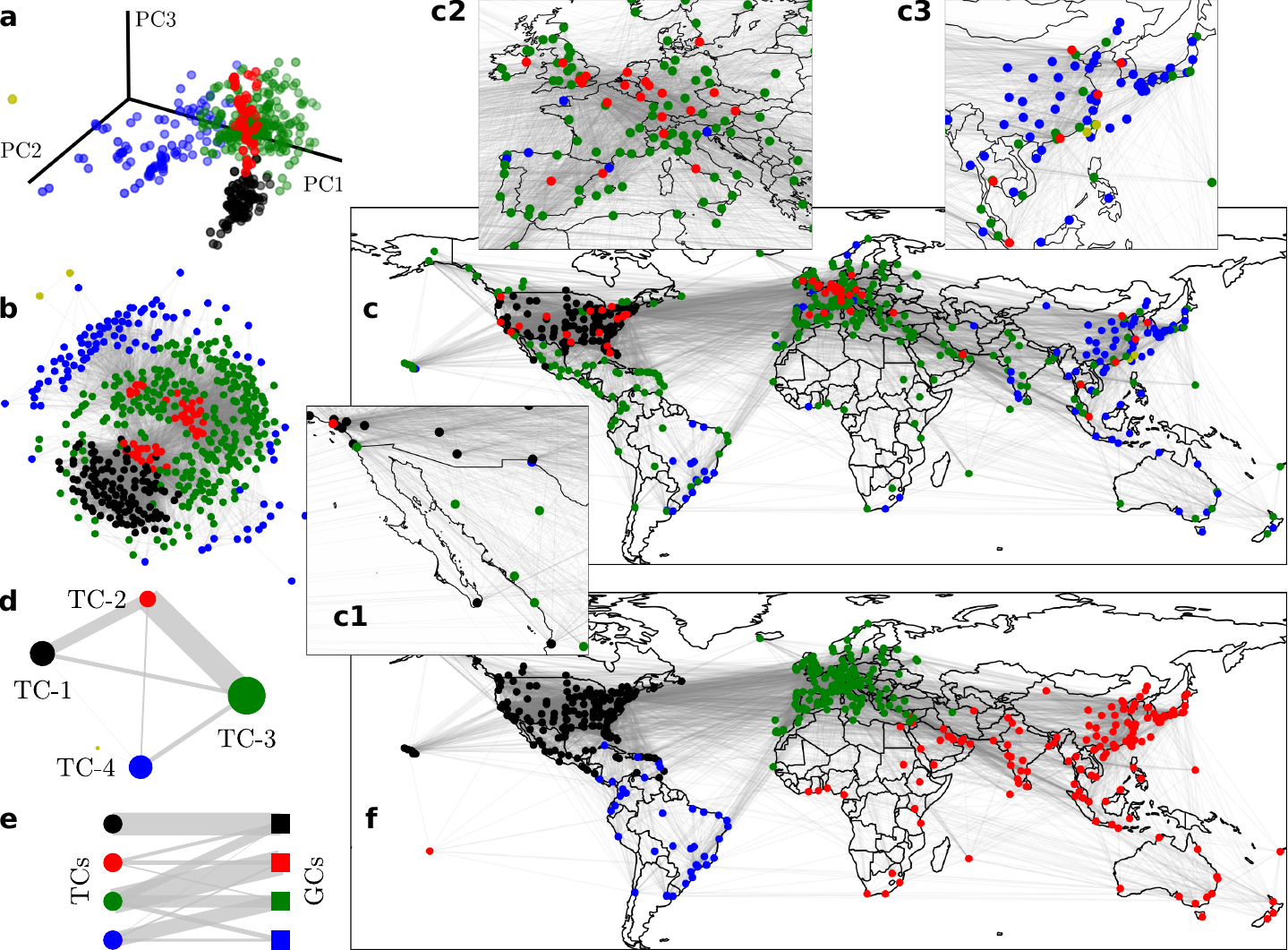}
  
      \caption{{\bf Topological Communities in the global transportation network.} {\bf a} Eigenspace of topological properties of airports as nodes of the global transportation network. Color-coding shows five TC (one in yellow, at the left, contains two oddball airports not discussed in the text). {\bf b} TC color projected onto the network in spring layout. {\bf c} Same color-coded TC projected on the world map (with regional details). {\bf d} Summary graph shows the relevance of the global backbone (red) through which most of the traffic flows. {\bf e} Bipartite network showing how many nodes of each TC belong to each of four classical Geometric Communities. {\bf f} GC projected on the world map, highlighting their geographic clustering which ignores different topological roles of nodes within a region. }
  
      \label{fig:2}
    \end{center}
  \end{figure*} 

  Airports make up a global transport network \cite{marcelino2012critical, networkResources}, $\mathcal{G}^{GTN}$ (App.\ \ref{app:0}). Its nodes spread over their PC eigenspace (Fig.\ \ref{fig:2}{\bf a}) differently to how $\mathcal{G}^{WS}$ and $\mathcal{G}^{CNB}$ vertices did on theirs. This anticipates a distinct decomposition. Some prominent TC appear distributed over network layout (Fig.\ \ref{fig:2}{\bf b}), as well as on a world map (Fig.\ \ref{fig:2}{\bf c}). TC-$1$ (black, which we term the US TC) contains most US airports except the major ones. TC-$2$ (red) contains the main hubs world-wide (including in the US). We name this the global backbone. Both the US TC and the global backbone score high in PC-$1$, which correlates positively with different node centralities (Sup.\ Fig.\ \ref{fig:SM.4}{\bf b}). These two TC are told apart along PC-$3$, which correlates, among others, with larger clustering and square clustering (Sup.\ Fig.\ \ref{fig:SM.4}{\bf d}). This indicates that, while both these TC are very central in the network, the global backbone has denser local connectivity, with each airport presenting more complete triangles among neighbors. Both TC-$3$ (green) and TC-$4$ (blue) contain medium-sized and smaller airports. TC-$3$ seems more Euro- and Caribbean-centered, while TC-$4$ clusters around South-East Asia and Brazil. But nodes of these TC appear all over the world (see details in Fig.\ \ref{fig:2}{\bf c})---their difference is not geographical, but topological. 

  The TC decomposition also aids in producing coarse-grained summary graphs (App.\ \ref{app:3}). Fig.\ \ref{fig:2}{\bf d} has condensed all nodes of each TC to show that the global backbone, despite containing less airports than each of the other three TC discussed, channels most of the connections, including from the US TC to the rest of the world. The global backbone only contains one node in the former Soviet block (PRG, Prague), and none in Africa, Latin America, India, Japan, or Australia. The US TC is quite self-contained---it is the only one fairly captured by classic GC (Fig.\ \ref{fig:2}{\bf e-f}). Hence, it consists of geographically and geometrically close airports, that are also topologically similar as graph nodes. The US TC's topology is also dissimilar to that of other world regions. This likely stems from the US historic decision to prioritize airborne transportation over, e.g., railway. TC insights can carry socioeconomic and strategic relevance. The airport most topologically similar to Mexico's Ciudad Ju\'arez (CJS) is Bod{\o} (BOO), in Norway. These might constitute the best models of each-other for planning logistics or expansions, even though they are an ocean and $4$ flights apart (compare to the graph's diameter, $5$, and average path length, $2.27$). 

  While one GC encloses the US TC completely, it also subsumes US airports from the global backbone. This misses a much more nuanced structure that is also erased as European, South American, and African-Asian-Austronesian nodes are grouped in their respective GC (Fig.\ \ref{fig:2}{\bf f}). Even though TC are geographically distributed, they remain fairly contiguous (Sup.\ Fig.\ \ref{fig:SM.11}{\bf a}). Notwithstanding, classic GC fail to capture them. Both TC and GC reveal key, complementary facets of complex networks. Neither approach is superfluous---both should be applied when we start studying a new graph. Geographical clustering of airports was appreciated in an earlier study \cite{guimera2005worldwide}, which also noted that, within each GC, nodes could play different roles. Hypothesis-driven measurements followed to clarify this (App.\ \ref{app:7}). The TC approach is more principled, incorporates more diverse measurements, and allows the graph to guide us towards its relevant internal structure.

  \begin{figure*}[] 
    \begin{center} 
      \includegraphics[width=\linewidth]{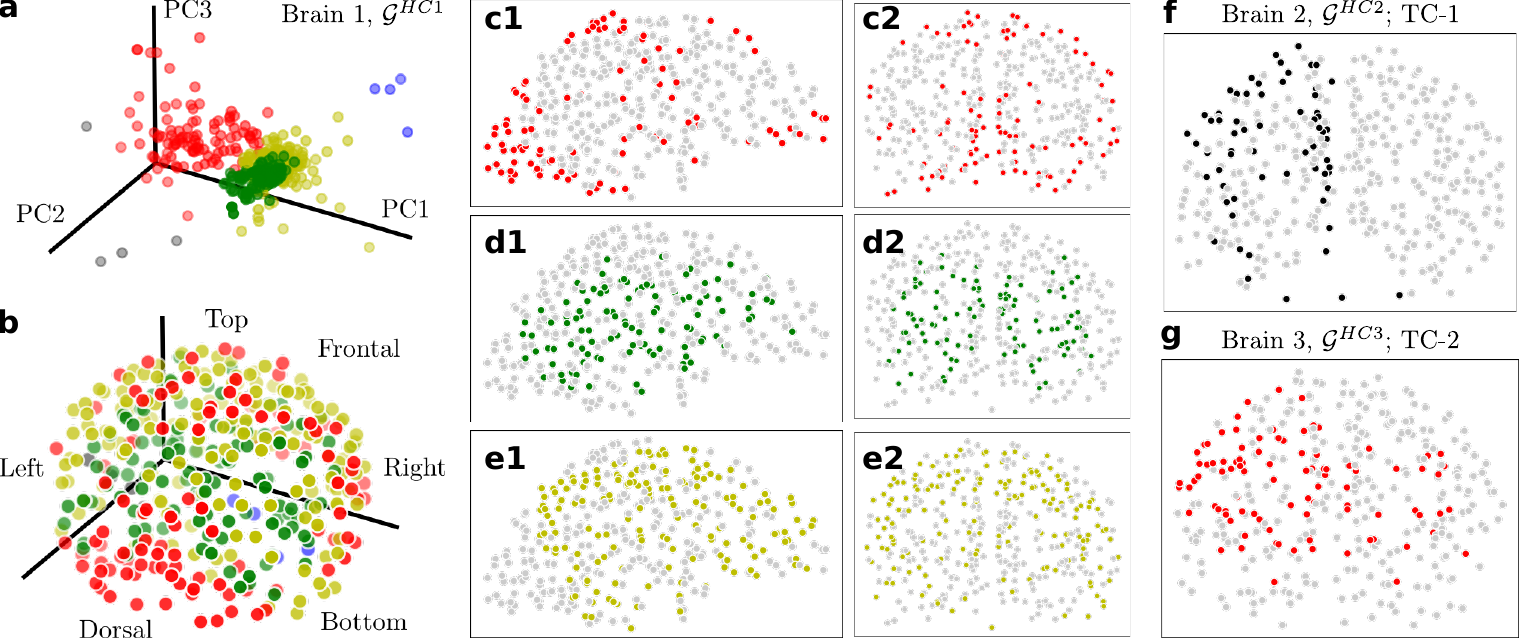}
  
      \caption{{\bf Topological Communities in human connectomes.} {\bf a} Eigenspace of topological properties for a brain, $\mathcal{G}^{HC1}$, with colors coding five TC (only three of which are noteworthy and discussed in the text). {\bf b} Color-coded TC projected on the brain. {\bf c-e} Sagittal and frontal views of each of the three salient TC from $\mathcal{G}^{HC1}$. {\bf c} TC-$2$, containing nodes from the visual and somatosensory cortex, among others. {\bf d} TC-$3$, containing mostly non-superficial nodes. {\bf e} TC-$5$, containing mostly superficial nodes. {\bf f} TC-$1$ of a different brain, $\mathcal{G}^{HC2}$, is highly asymmetric, containing mostly nodes from the left hemisphere. {\bf g} TC-$2$ of yet another brain, $\mathcal{G}^{HC3}$, is asymmetric in a different way. }
  
      \label{fig:3}
    \end{center}
  \end{figure*} 

  Finally, let us turn our attention to connectomes, which summarize connectivity patterns in the brain. We downloaded $1,064$ connectomes in which nodes are small brain volumes and an edge exists if at least one axonal fiber was detected connecting the corresponding volumes \cite{humanConnectome, kerepesi2017braingraph} (App.\ \ref{app:0}). Exhaustive analysis will follow in future works. Here we intend to illustrate TC, for which we focus on three brains: $\mathcal{G}^{HC1}$, $\mathcal{G}^{HC2}$, and $\mathcal{G}^{HC3}$. $\mathcal{G}^{HC1}$ was chosen because, after visual inspection of numerous connectomes, it presents three TC that are clear-cut and common to many other brains. $\mathcal{G}^{HC1}$ nodes spread over PC-eigenspace again differently to previous examples (Fig.\ \ref{fig:3}{\bf a}), suggesting a novel decomposition. The most outstanding cluster (TC-$2$, red) includes nodes from the somatosensory and primary visual cortices (Fig.\ \ref{fig:3}{\bf b-c}). This is perhaps the feature that we observe more often across the database. TC-$2$ suggests that the striate and somatosensory cortices are topologically similar (even though they are non-contiguous) and singularly distinct from other brain regions. TC-$3$ (green) contains mostly nodes located deeper in the brain, while TC-$5$ (yellow) is more superficial (Fig.\ \ref{fig:3}{\bf b}, {\bf d-e}). This might seem a trivial division; but it further highlights TC-$2$, which consists of mostly superficial nodes, yet differently grouped than other cortical regions. Also, the deep-superficial divide varies across brains in the database. 

  Connectomes $\mathcal{G}^{HC2}$ and $\mathcal{G}^{HC3}$ were chosen to illustrate symmetry breaking in the brain. In $\mathcal{G}^{HC2}$, its TC-$1$ (black, Fig.\ \ref{fig:3}{\bf f}) groups mostly superficial, but only left-hemispheric, nodes. Compare this with the largely symmetric $\mathcal{G}^{HC1}$, for which TC spanned both hemispheres---mirror symmetric, yet distant nodes had a similar topology. But for $\mathcal{G}^{HC2}$, certain left-hemispheric nodes are topologically more similar to each other than to their mirror-symmetric counterparts. $\mathcal{G}^{HC3}$ presents a more complicated decomposition. Its TC-$2$ (red, Fig.\ \ref{fig:3}{\bf g}) contains superficial nodes in the left hemisphere but deeper ones at the right. Symmetry and symmetry breaking are prominent brain features with clinical implications \cite{davidson1995brain, seoane2020modeling, carballo2022phase}, often linked to optimality and complexity \cite{seoane2020fate, seoane2023optimality}. But they are not easy to formalize and measure (recent efforts mobilized consortia with hundreds of researchers \cite{kong2014mapping,kong2022mapping}). TC nimbly report topological symmetry and asymmetry---a research line that we will explore in the future. TC decomposition offers a refined structural analysis complementary to GC, which again fail to recover outstanding topological features (Sup.\ Fig.\ \ref{fig:SM.19}). Cortical centers singled out by TC participate of distinct cognitive processes. Our analysis suggests a principled way to further explore the effect of connectome topology on cognition by correlating TC and functional regions---whcih we will also explore in the future. 

  App.\ \ref{app:6} showcases brief analyses of networks of programming languages \cite{valverde2015punctuated}, a macaque connectome \cite{harriger2012rich}, yeast protein-protein interactions \cite{michaelis2023social}, and bill co-sponsorship in the US house of representatives \cite{neal2014backbone, andris2015rise, neal2020sign, hohmann2023quantifying}. These studies illustrate a range of TC decompositions, highlighting the many ways in which a few topological building blocks can be arranged to produce complex networks. Some novel insights hinted at in App.\ \ref{app:6} will be developed in dedicated papers. The wealth of additional data where to apply this framework is vast. Our analysis is inspired by earlier observations that topologial roles might vary within classic GC \cite{guimera2005worldwide}, as well as by methods of numerical topology and dimensionality reduction in vogue in computational neuroscience \cite{gallego2017neural, houston2022squishy, gardner2022toroidal, sebastian2023topological} or cell biology \cite{becht2019dimensionality}. We dicuss connections with earlier work in App.\ \ref{app:7}. TC offer a novel network decomposition perhaps on par in importance with classic community detection---itself a cornerstone of network science. $20$ years after their introduction \cite{fortunato202220}, GC are a vibrant research field both for the development of more refined algorithms \cite{palla2005uncovering, bianconi2013statistical, peixoto2022disentangling} and as a revealing tool across the sciences \cite{palla2005uncovering, van2012high, neal2014backbone, andris2015rise, neal2020sign, hohmann2023quantifying}. We expect similar applicability for TC. This paper aims at introducing the framework, but tweaks and refinements should expand its possibilities. We limited ourselves to unweighted, undirected graphs---both for conceptual simplicity and ease to handle topological properties. As we introduce directedness and weights, we expect more distinct TC to appear---i.e.\ more topological building blocks available that can be arranged in more different ways, resulting in new insights across networks. An enticing application might help open the black box of Artificial Intelligence by studying TC in Artificial Neural Networks. The paradigm should also work on multiplex graphs \cite{mucha2010community, nicosia2013growing, bianconi2013statistical} or simplicial networks \cite{moore2012analyzing, patania2017shape}. On the technical side, we used PC again for simplicity, but a range of dimensionality reduction methods could be applied instead \cite{kramer1991nonlinear, mcinnes2018umap, van2008visualizing}. What matters is the TC conceptual framework, which we think offers a new, relevant tool in network science, covering a blind spot in graph analysis. A researcher who is confronted with a new network and asks ``What can this graph tell me? What kind of analyses should I run on it?'', should, by default, try out GC and TC to let the graph shine a light on its most salient geometric and topological features. 

\vspace{0.2 cm}

  \section*{Acknowledgments}

  	The author thanks Susanna Manrubia for her support and for early feedback provided about this manuscript (together with Iker Atienza), and for her coming up with the term {\em Topological Communities}, for which I am deeply indebted. The author also acknowledges the help and support of the ``extended'' group of Susanna Manrubia and Jose Cuesta, whose biweekly seminars helped shape this paper. This work has been supported by the Occident Foundation (grant FJSCNB-2022-12-B), and by the Spanish State Research Agency, AEI, through the ``Severo Ochoa'' Programme for Centres of Excellence in R\&D (CEX2023-001386-S) and, together with the Spanish Department for Science and Innovation (MICINN), through project PID2023-153225NA-I00 within the State Program for Knowledge Generation.

\appendix 

  \setcounter{figure}{0}
  \setcounter{table}{0}
  \setcounter{page}{1}
  \renewcommand{\figurename}{SUP. FIG.}

    \section{Networks}
      \label{app:0}

      \begin{table}[h] 
        \begin{tabular}{@{}lcccc@{}}
	        \hline
	        Network & Symbol & $N^n$ & $N^e$ & References \\
	        \hline
	        Random Watts-Strogatz & $\mathcal{G}^{WS}$ & $300$ & $600$ & \cite{watts1998collective} \\
	        CNB collaboration network & $\mathcal{G}^{CNB}$ & $286$ & $906$ & \cite{manrubia2022report} \\
	        Top $500$ global transport network & $\mathcal{G}^{GTN}$ & $500$ & $13038$ & \cite{marcelino2012critical, networkResources} \\
	        MRI human connectomes & $\mathcal{G}^{HC1}$ & $384$ & $7668$ & \cite{kerepesi2017braingraph} \\ 
	        MRI human connectomes & $\mathcal{G}^{HC2}$ & $369$ & $6573$ & \cite{kerepesi2017braingraph} \\ 
	        MRI human connectomes & $\mathcal{G}^{HC3}$ & $353$ & $5641$ & \cite{kerepesi2017braingraph} \\ 
	        \hline
	        Programing languages & $\mathcal{G}^{PL}$ & $353$ & $5641$ & \cite{valverde2015punctuated} \\ 
	        Macaque connectome & $\mathcal{G}^{MC}$ & $242$ & $3054$ & \cite{harriger2012rich} \\ 
	        Yeast protein-protein & $\mathcal{G}^{Y}$ & $3839$ & $30955$ & \cite{michaelis2023social} \\ 
	        US House bill co-sponsorship & $\mathcal{G}^{US93}$ & $433$ & $8006$ & \cite{neal2014backbone, neal2020sign, neal42constructing} \\ 
	        US House bill co-sponsorship & $\mathcal{G}^{US114}$ & $434$ & $26576$ & \cite{neal2014backbone, neal2020sign, neal42constructing} \\ 
	        \hline
        \end{tabular}

        \caption{{\bf Case studies to introduce topological communities}. The first block of networks is discussed in the main text and in more detail in the appendixes. The second block of networks is briefly discussed in App.\ \ref{app:6} to showcase the diversity of TC decompositions and the range of applicability of this paradigm. } 
        \label{tab:A0} 
      \end{table}

      A graph or network, $\mathcal{G}$, consists of a set, $V$, containing $N^n \equiv |V|$ nodes or vertices $v_i \in V$; and a set, $E$, containing $N^e \equiv |E|$ unordered tuples of the kind $(v_i, v_j) \in E$ that indicate that nodes $v_i$ and $v_j$ are connected. We call the elements of $E$ edges or links indistinctly. We will say $(v_i, v_j) \in E$ if either $(v_i, v_j) \in E$ or $(v_j, v_i) \in E$. It is convenient to introduce the adjacency matrix: $A \equiv \{a_{ij};\>\>i=1, \dots, N^n;\>j=1, \dots, N^n \}$ with $a_{ij}=1$ if $(v_i, v_j) \in E$. It is also convenient to introduce the neighborhood of a node: $V_i \equiv \{v_j \in V, (v_i, v_j) \in E \}$. In this paper we only work with unweighted, undirected graphs with no self-loops (hence $(v_i, v_i) \notin E$ for any $i$). 

      All networks studied in this paper are unweighted and undirected. We will extend our methods to weighted and directed networks in following papers. We summarize some characteristic of our case-study graphs in table \ref{tab:A0}. An itemized list follows with additional details where necessary: 
      \begin{itemize}

        \item Random Watts-Strogatz (WS) network, $\mathcal{G}^{WS}$: We generated a random WS network \cite{watts1998collective} with $300$ nodes, connecting each vertex to its 4 nearest neighbors. The rewiring probability was $0.05$. 

        \item CNB collaboration network, $\mathcal{G}^{CNB}$: In \cite{manrubia2022report}, we built the collaboration network of researchers at the Author's home institution, the Spanish National Centre for Biotechnology (CNB). We focused on the time period $2016-2021$, during which the CNB was distinguished as a Severo Ochoa center of excellence by the Spanish Ministry for Science, Innovation, and Universities. We collected all papers published by CNB researchers during that period. Each CNB researcher constitutes a node in the network, and two vertices are connected if the corresponding scientists coauthored at least one paper within the studied period. Edges leaving the graph (i.e.\ collaborations with external researchers) are ignored. 

        \item Top $500$ global transport network, $\mathcal{G}^{GTN}$: We used data from \cite{marcelino2012critical} (available at \cite{networkResources}) to recreate the global network that connects every two airports between which there is at least one flight. This data is restricted to the top $500$ airports in volume of passengers. 

        \item MRI human connectomes, $\mathcal{G}^{HC}$: Networks downloaded from \cite{humanConnectome}, generated by \cite{kerepesi2017braingraph}. The dataset contains several libraries of connectomes generated from MRI from the Human Connectome Project \cite{humanConnectomeProject}. In the original dataset, the brain has been divided into voxels. We are given the number of fibers connecting any two brain regions, which were inferred from diffusion MRI images using standard techniques \cite{kerepesi2017braingraph}. We assume that two nodes are connected if at least a fiber exists between the corresponding regions. The dataset contains $1,064$ brains with $463$ regions in each connectome (note that we work with the largest connected components, so the final number of nodes varies from one brain to another). In this paper we discuss connectomes ob subjects $101309$ ($\mathcal{G}^{HC1}$), $992774$ ($\mathcal{G}^{HC2}$), and $989987$ ($\mathcal{G}^{HC3}$) within the Human Brain Connectome database. A more thorough analysis will be presented in successive papers. 

        \item Programming languages, $\mathcal{G}^{PL}$: In \cite{valverde2015punctuated}, a phylogenetic tree of programming languages was built based on which got inspiration from each-other (as documented in each language's manual). This is a directed network, but we reduced the graph to its undirected version. 

        \item Macaque connectome, $\mathcal{G}^{MC}$: We used the network in \cite{harriger2012rich}, which built upon collated data from $410$ tract tracing studies found in the CoCoMac database (http://cocomac.org; http://cocomac.g-node.org; \cite{stephan2001advanced, kotter2004online}). We ignored direction and weights of the connections. Opposed to our human connectomes, this network does not correspond to a unique brain, but to the result of merging data from several macaques. 

        \item Yeast protein-protein interaction network, $\mathcal{G}^{Y}$: We used the most recently published data on protein-protein interaction in yeast {\em Saccharomyces cerevisiae} \cite{michaelis2023social}. While data in \cite{michaelis2023social} is provided as directed links, this network is naturally undirected (interaction of protein A with B implies an equal interaction of B with A). Data is also unweighted. 

        \item Bill co-sponsorship network within the US house, $\mathcal{G}^{US}$: We used the tools developed in \cite{neal2014backbone, neal2020sign, neal42constructing} and available in \cite{incidentally} as an R package. These tools allow us to reconstruct co-sponsorship networks within the US house and senate. Representatives in either chamber can support bills introduced for consideration. The software in \cite{incidentally} builds a network considering whether two representatives co-sponsor bills together more often than by random chance. The resulting network is unweighted and undirected. Available data spans from the $93$rd to the $114$th congresses. We discuss only two networks from the house---the first and last congresses, noted $\mathcal{G}^{US93}$ and $\mathcal{G}^{US114}$ respectively. Study of the complete dataset is left for future work. 

      \end{itemize}

    \section{Topological properties}
      \label{app:1}

      \begin{table}[h] 
        \begin{tabular}{@{}lcc@{}}
	        \hline
	        Property name & Formula & References \\
	        \hline
	        Degree & $k_i = |V_i|$ & --- \\
	        Eigenvector centrality & $C^E_i \equiv \nu^1_i$ &  \cite{bonacich1987power, newman2018networks} \\
	        Betweenness centrality & $C^B_i \equiv \sum_{j,k} \sigma(j,k|v) / \sigma(j,k)$ & \cite{brandes2001faster} \\
	        Closeness centrality & $C^C_i \equiv (N_n-1)/\sum_{j}d(i,j)$ & \cite{freeman2002centrality}\\
	        Harmonic centrality & $C^H_i \equiv \sum_{j\ne i} 1/d(i,j) $ & \cite{boldi2014axioms}\\
	        PageRank centrality & $C^P_i$ & --- \\
	        Coreness or core number & $\kappa_i \equiv \max_k\{v_i \in k\text{-core}\}$ & \cite{batagelj2003m} \\
	        Onion layer & $\mathcal{L}_i$ & \cite{hebert2016multi, allard2019percolation}\\
	        Effective size & $E_i \equiv k_i - 2t_i/k_i$ & \cite{lazega1995structural, borgatti1997structural} \\
	        Clique number & $\omega_i$ & --- \\
	        Number of cliques & $N^\omega_i$ & --- \\
	        Number of triangles & $N^t_i \equiv \sum_{j,k \in V_i} a_{jk}$ & --- \\
	        Cycle ratio & $r_i$ & \cite{fan2021characterizing} \\
	        Number of minimum cycles & $N^c$ & \cite{fan2021characterizing} \\
	        Inverse of max.\ min. cycle & $\mu$ & \cite{fan2021characterizing} \\
	        Inverse of node girth & $\gamma$ & \cite{fan2021characterizing} \\
	        Clustering coefficient & $C^3_i \equiv N^t_i/k_i(k_i-1)$ & \cite{newman2018networks} \\
	        Square clustering coefficient & $C^4_i$ & \cite{lind2005cycles, zhang2008clustering}\\
	        Node constraint & $L_i \equiv \sum_{j\in V_i} l(i,j)$ & \cite{burt2004structural} \\
	        \hline
        \end{tabular}

        \caption{{\bf Primary topological properties employed in our analysis}. For constraint, $L_i$, the quantity $l(i,j)$ sstands for {\em local constraint} and it is computed as $l(i,j) \equiv \left( a_{ij} + \sum_{k} a_{ik}a_{jk}\right)^2$. }  
        \label{tab:A1} 
      \end{table}

      For our topological analysis, we measure a series of properties for each node. We first chose a set of {\em primary properties} (Tab.\ \ref{tab:A1}) with the hope that they describe all relevant topological aspects of a node exhaustively. Specifically, we try to capture dimensions such as centrality (which can be of different kinds---e.g.\ eigenvector, betweenness, etc.), density of local connections (as measured by cliques and $k$-cores), edges (girth or abundance of minimal cycles associated to a node), or overlap between neighbor connections (effective size or constraint). In this appendix we define in detail the properties just mentioned and others. 

      Some of these properties can carry similar information as others. Which measurements are redundant usually changes from one network to another---hence each graph induces a similarity structure between node properties (see App.\ \ref{app:2}). Adequate dimensionality reduction methods prevent redundancies from biasing our results (see App.\ \ref{app:2}). Important topological aspects might have been left out despite our efforts. This could be alleviated in the future by introducing additional measurements. Our central contribution in this paper is only contingent on these details. We have summarized the chosen primary properties, their formulas or notation, and some useful references in Tab.\ \ref{tab:A1}. Below we expand these properties in an itemized list with lengthier explanations where needed. All numerical evaluations in this paper have been implemented in Python using NetworkX \cite{hagberg2008exploring}. 

      Let us take an arbitrary primary property, $\pi$, and note $\pi^\mathcal{G} \equiv \{ \pi_i, v_i \in \mathcal{G} \}$ the result of numerically evaluating this quantity over all nodes in network $\mathcal{G}$. For each primary property we derive two additional {\em secondary properties}: (i) the average over a node's neighbor, $\left<\pi\right>_i \equiv \sum_{j\in V_i} \pi_j/k_i$, where $k_i$ is the node's degree; and (ii) the standard deviation over a node's neighbor, $(\pi)_i \equiv \sqrt{ \sum_{j\in V_i} (\pi_j - \left<\pi\right>_i)^2/k_i }$. We can run our analysis including or excluding secondary properties---actually, we can run it excluding any combination of measurements, also primary ones. We have found that including secondary properties enriches our methods, suggesting that they capture salient information about each node, and that this allows grouping up vertices that have similar relationships to their neighbors even if they are not contiguous in the network.  

      \begin{figure*}[] 
        \begin{center} 
          \includegraphics[width=\textwidth]{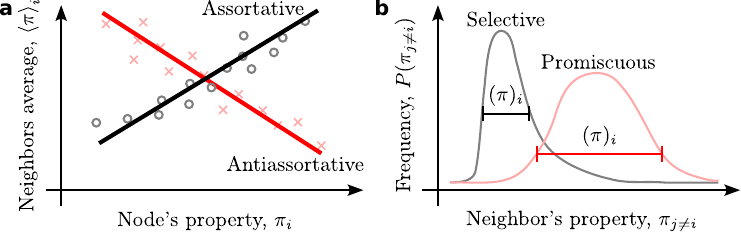}
      
          \caption{{\bf Some information that might be revealed by secondary properties.} {\bf a} Average measurements over a node, when compared with the node's properties, tell us whether a vertex tends to link to other similar or dissimilar ones. {\bf b} Standard deviation of measurements over a node's neighbors tells us whether a vertex is selective or promiscuous in choosing its neighborhood. }
      
          \label{fig:SM.1}
        \end{center}
      \end{figure*}

      %
      %
      The first set of secondary properties, $\left<\pi\right>_i$, can tell us whether nodes tend to connect with vertices which are similar or dissimilar to themselves (Sup.\ Fig.\ \ref{fig:SM.1}{\bf a}). This will result in correlations or anti-correlations during PCA, which allows us to generalize ideas of assortativity. Assortativity is used to indicate that nodes with a high degree are connected to others with high degree as well. Assortative networks emerge spontaneously from entropic forces alone---given a configuration, they are much more common \cite{johnson2010entropic}. In anti-assortative graphs, high-degree nodes avoid each-other and prefer to link with less-connected vertices. This is rarer, suggesting specific mechanisms operating in that direction. Examples of antiassortative graphs are syntax networks or genotype networks explored by viruses. We do not need to stop at node degree. Given a network, do those with large betweenness centrality tend to connect with others scoring also high in this quantity? What about the number of cycles that a node is involved in? If such trends are relevant for some property in a network, our analysis will pick them up. 

      The second set of derived properties, $(\pi)_i$, tells us whether a node is picky regarding which other vertices it connects to (Sup.\ Fig.\ \ref{fig:SM.1}{\bf b}). If $(\pi)_i$ is small, then $v_i$ tends to connect with others within a specific range of values of property $\pi$. If $(\pi)_i$ is larger, the neighbors of $v_i$ are heterogeneous. What is small or large only makes sense within the context of the whole network. Again, our analysis provides an automatic way to report on this aspect if it is salient in the graph. 

      \subsection{Itemized list of primary properties}
        \label{app:1.1}

        \begin{itemize}

          \item Node degree: $k_i \equiv |V_i|$. 

          \item Eigenvector centrality \cite{bonacich1987power, newman2018networks}: Let $\nu^1$ be the eigenvector that corresponds to the largest eigenvalue of the Adjacency matrix. Then, $\nu^1_i$ is the $i$-th entry of this eigenvector, and it corresponds to the $i$-th node eigenvector centrality. 

          \item Betweenness centrality \cite{brandes2001faster, newman2018networks}: Let $\sigma(j,k)$ be the number of shortest paths connecting $\nu_j$ and $\nu_k$. Let $\sigma(j,k|i)$ the number of such shortest paths that pass through $\nu_i$. Then the betweenness centrality reads: $C^B_i \equiv \sum_{j,k} \sigma(j,k|i) / \sigma(j,k)$. 

          \item Closeness centrality \cite{freeman2002centrality}: This property measures the inverse of the average closest distance of a node to all others. Let $d(i,j)$ be the shortest distance between $\nu_i$ and $\nu_j$. Then $C^C_i \equiv (N_n-1)/\sum_{j\ne i}d(i,j)$. 

          \item Harmonic centrality \cite{boldi2014axioms}: Related to the previous one, this quantity measures the average of the inverse of closes distances of a node to all others: $C^H_i \equiv \sum_{j\ne i} 1/d(i,j)$. 

          \item PageRank: Pagerank is a popular algorithm that ranks nodes from most to least central (in the eigenvector centrality sense). It obviously correlates with eigenvector centrality, but it is non-linearly related to it and rather provides information about cumulative centrality (much as a cumulative distribution relates to a density distribution in statistics). 

          \item Coreness or core number \cite{batagelj2003m}: A $k$-core is found by iteratively removing all nodes with degree less than $k$ until no more nodes can be removed. The coreness or core number of a node is the largest $k$-core to which a node belongs. 

          \item Onion layer \cite{hebert2016multi, allard2019percolation}: In the iterative process to compute a $k$-core we remove nodes sequentially. Assuming a connected component, to find the $2$-core we first remove nodes with degree $1$, as they have less than $k=2$ connections. These nodes belong to the most external onion layer. If, after removing this layer, we are left only with nodes with degree equal or larger than $2$, we have found the $2$-core (which might consist of a connected graph or many). Otherwise, after removing the first layer, a new set of nodes will be left with degree less than $2$. This is the second onion layer. We remove them and repeat the process until the $2$-core is located. Note again that this might be a unique connected component or many. Next we set up to find the $3$-core, which is contained within the $2$-core; then proceed for higher $k$-cores until none is found. A node's onion layer is the order in which it is removed in this process. 

          \item Effective size \cite{lazega1995structural}: The ego-network of node $v_i$ (named ego-node in this context), $\mathcal{G}_i$, is the subgraph $\mathcal{G}_i \subset \mathcal{G}$ that contains all neighbors of $v_i$. If nodes within an ego-network are linked, these connections are redundant in a very specific sense---e.g.\ because information will arrive repeatedly through many paths. {\em Effective size} is an attempt to capture this redundancy. In undirected, unweighted neighbors, it is straightforwardly $E_i \equiv k_i - 2t_i/k_i$ \cite{borgatti1997structural}. , where $k_i$ is the ego-node's degree and $t_i$ is the number of edges within $\mathcal{G}_i$ that do not involve $v_i$. 

          \item Clique number: A clique is a graph in which all nodes are connected to each-other. A node's clique number is the size of the largest clique to which it belongs within the larger network, $\mathcal{G}$. Mathematically, over all subgraphs $g$ within a network $\mathcal{G}$, $\omega_i \equiv \max_\omega \{ \omega \equiv | g \subset \mathcal{G} |;\>\> v_i \in g\>\> \wedge \>\> g\>\>\text{is clique} \}$. 

          \item Number of cliques: Number of maximal cliques that a node belongs to. Mathematically, $N^\omega \equiv |\{ g\subset \mathcal{G},\>\> v_i \in g\>\>  \land |g|=\omega_i\>\> \land g\>\> \text{is clique}\}|$. 

          \item Number of triangles: Given a node, $v_i$, and its closest neighbors, $V_i$, a triangle is completed if $v_j \in V_i$ and $v_k \in V_i$ and $(v_k, v_j) \in E$. Thus, $N^t_i \equiv \sum_{j,k \in V_i} a_{jk}$. This property is tightly related to clustering. In our analysis we found that, in most networks, nodes with a large centrality also showed a very small clustering. This is so mostly because a node with a large centrality has got many more potential triangles and it is much more difficult that it will complete them all. We speculated that nodes with small degree might present high clustering even though they had a small number of associated triangles, hence that this quantity behaves differently to clustering and that it might be informative in some networks. 

          \item Cycles or loops: Cycles are topologically relevant features. Characterizing a graph's cycle structure is particularly difficult because the number of loops grows combinatorially with network size. We tested several options in smaller graphs before deciding on the four properties reported next. One option was to produce a cycle basis---a set of loops from which all others in a graph can be generated---but each network comprises combinatorially many different bases. A stochastic evaluation was a possibility. Another option was to retain the so-called minimal basis, but finding it grows faster than polynomially with network size. We think that there is much room for improvement in the characterization of a graph's cycle structure. Whenever new breakthroughs appear, they can be seamlessly incorporated into our analysis. Here we opted to use recently published work \cite{fan2021characterizing} that localizes, for each node, an associated set of minimal cycles, $S_i  \equiv \{ \sigma^l_i, l=1, \cdots, N^c_i \}$. Here, $N^c_i$ is the number of minimal cycles associated to the vertex $v_i$ and each $\sigma^l_i$ is a collection of vertices $\sigma^l_i \equiv \{ v^l_i(1), \dots, v^l_i(\lambda^l_i) \}$ where $\lambda^l_i \equiv |\sigma^l_i|$ is the length of the cycle. $\sigma^l_i$ are such that an edge exists in $\mathcal{G}$ connecting each two consecutive nodes in $\sigma^l_i$, $(v^l_i(m), v^l_i(m+1)) \in E$, and the last and first nodes of $\sigma^l_i$, $(v^l_i,(\lambda^l_i), v^l_i(1)) \in E$. See \cite{fan2021characterizing} for more details. From this set, we compute: 
            \begin{itemize}

              \item Cycle ratio: Following \cite{fan2021characterizing}, from the above set we compute the matrix entries $c_{ij}$ as the number of loops in $\cup_i S_i$ that contain both vertices $v_i$ and $v_j$. The cycle ratio is defined as $r_i = 0$ if $v_i$ has no cycles associated and $\sum_j {c_{ij}/c_{ii}}$ otherwise. This measures the presence of node $v_i$ in the loops associated to other vertices. 

              \item Number of minimum cycles: We score the size of the set of minimum cycles associated to each node, $N^c_i$. 

              \item Inverse of maximum minimum cycle: Among a node's minimum cycles, there is one (or many) with largest length. We wanted to include this information in the analysis, but nodes without associated loops were troublesome. A possibility was to assign them a maximum minimum cycle of $0$, but this would introduce an artificial proximity to vertices with small associated cycles. We opted for assigning an infinity-length cycle to such nodes, then working with the inverse of this quantity to avoid numerical problems. Thus, $\mu_i \equiv 1/max_{|\sigma^l_i|}\{S_i\}$. 

              \item Inverse of node girth: This measure presented the same problem as the previous one, and it was solved with the same strategy. A node's girth is the size of the smallest associated cycle. We take its inverse: $\gamma \equiv 1/min_{\lambda^l_i}\{S_i\}$. 

            \end{itemize}

          \item Clustering coefficient: fraction of possible triangles through a node that exist, $c_i \equiv N^t_i/k_i(k_i-1)$. 

          \item Square clustering coefficient: fraction of possible squares involving a node that exist. This property was developed to attempt a kind of clustering coefficient for bipartite networks, in which triangles are never possible \cite{lind2005cycles, zhang2008clustering}. 

          \item Constraint: Node constraint is an alternative way to tackle the redundancy of connections within the immediate neighborhood of a node \cite{burt2004structural}. It is a measurement introduced in economics to quantify how much investment overlap there is between neighbor nodes. 

        \end{itemize}

    \section{Principal components analysis}
      \label{app:2}

      Some of the measurements might have taken infinite values, or might be the same for all nodes. In this last case, they do not offer any relevant information that clarifies variety of node topology within the graph. We detect and remove these pathological properties before our analysis. We are left with an array, $\Pi_i \equiv \{\pi^l_i,\> l=1, \dots, N^p\}$, where $N^p$ is the total number of properties of interest. 

      $\Pi \equiv \{\Pi_i,\> i=1, \dots, N^n\}$ contains valuable information about our network, $\mathcal{G}$. Analyses in network science are often driven either by guesses after visual inspection (e.g.\ because a community structure is outstanding, even though this can be deceiving \cite{peixoto2021descriptive}) or hypothesis validation (e.g.\ we want to check out whether our graph is assortative, whether it is a small world, etc.). Instead, our analysis asks the network to guide us towards its relevant features. Which facets are important usually changes from one network to another. For this task we can use any available dimensionality reduction techniques---e.g.\ autoencoders \cite{kramer1991nonlinear}, umap \cite{mcinnes2018umap}, or other non-linear manifold embeddings \cite{van2008visualizing}. For simplicity, we choose the most straightforward one, Principal Component Analysis (PCA) \cite{pearson1901liii}. This also allows a more intuitive discussion that helps us focus on the novelty of Topological Communities, not on technicalities. More modern methods will doubtlessly enrich our framework in the future. 

      \begin{figure*}[] 
        \begin{center} 
          \includegraphics[width=\linewidth]{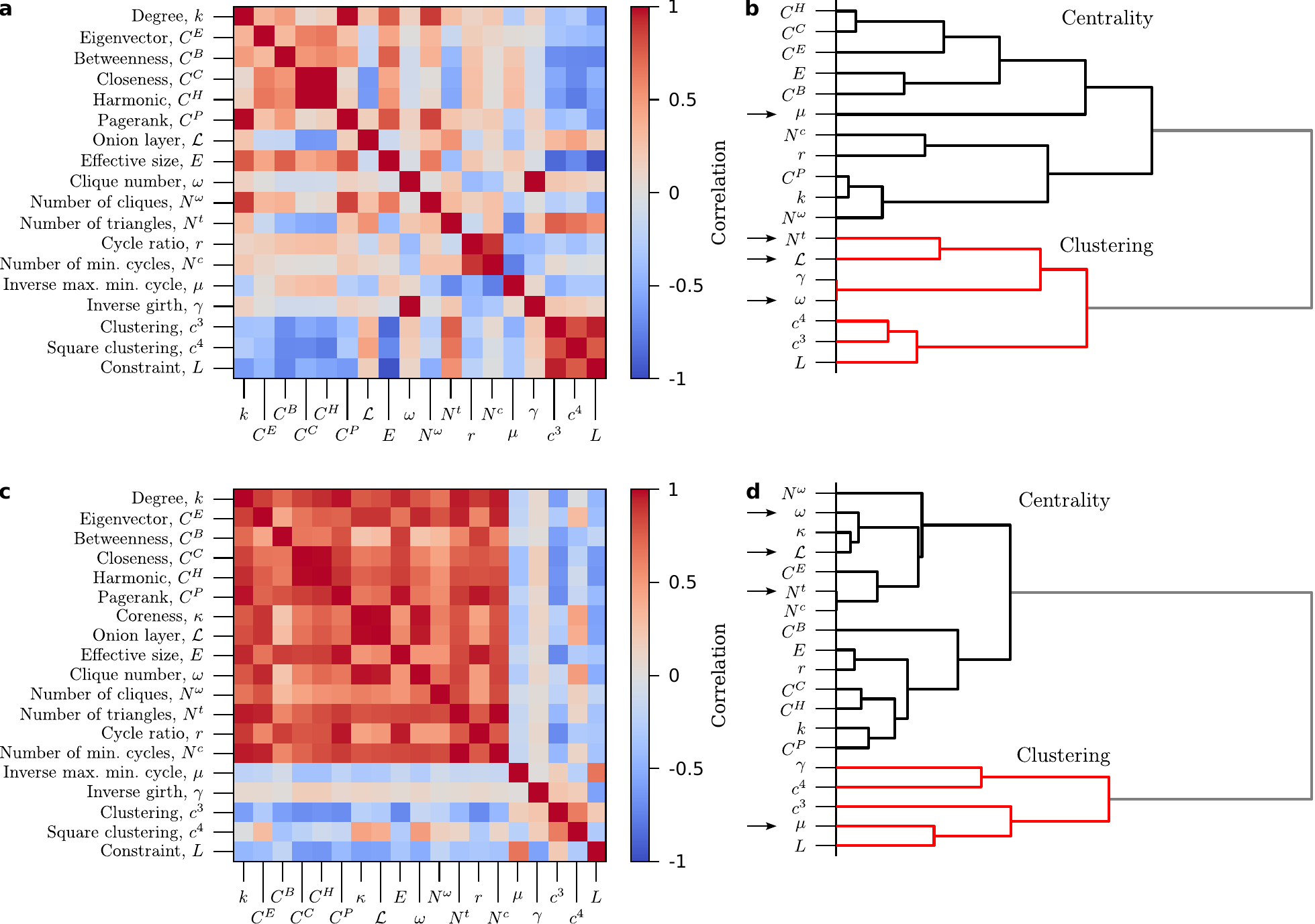}
      
          \caption{{\bf Each network induces a different correlation structure between properties.} {\bf a} Correlation matrix for primary properties of a Watts-Strogatz network. {\bf b} Dendrogram illustrating which measurements capture more similar structure than each-other in the Watts-Strogatz network. {\bf c} Correlation matrix for primary properties of the global airport network. {\bf d} Dendrogram depicting similarity between properties in the global airport network. }
      
          \label{fig:SM.2}
        \end{center}
      \end{figure*}

      We center and normalize all variables before computing the correlation matrix. These matrices start showing us important information about global topological properties in each graph. Sup.\ Fig.\ \ref{fig:SM.2}{\bf a} and {\bf c} show, respectively, correlation matrices for our WS graph, $\mathcal{G}^{WS}$, and the global airport network, $\mathcal{G}^{GTN}$. In this example we only show primary properties to simplify our visualizations. Each network induces a different correlation structure between our primary properties. To the question: ``Given a graph, do two distinct properties measure a same thing?'' The answer is: ``It depends on the specific network that we are looking at.'' For example, the onion layer decomposition of a node most often correlates with that vertex' centrality and degree, so we might think that these quantities are always similar. But above we showed that this is not the case for a WS network, where nodes with high centrality might be in low onion layers. 

      From the correlation matrices we apply hierarchical clustering (using the fcluster tool from SciPy) to derive dendrograms (Sup.\ Fig.\ \ref{fig:SM.2}{\bf b} and {\bf d}) that summarize which properties are more similar to each other in a given network. Across all networks studied we tend to observe two blocks: Those that correlate with centrality (black branches) and those that correlate with clustering (red). This split is not always well defined (see the WS graph in Sup.\ Fig.\ \ref{fig:SM.2}{\bf a-b}). When they are clear, these blocks are usually anti-correlated with each other. Clustering coefficient usually anti-correlates with centrality because very central nodes tend to have many more nearest neighbors and it is hence more difficult to complete all possible triangles. However, this relationship if far from trivial or parsimoniously lineal, we think that it deserves further study. 

      In the figure, arrows show properties that switch blocks when moving from $\mathcal{G}^{WS}$ to $\mathcal{G}^{GTN}$. This shows that the information that a property contains is contingent on the graph and must be understood in relation to other quantities. Take the onion layer, $\mathcal{L}$. The way that this property has been built \cite{hebert2016multi, allard2019percolation}, we would think that central nodes would be removed last, thus have a higher $\mathcal{L}$. This is the case in $\mathcal{G}^{GTN}$, but not in $\mathcal{G}^{WS}$. WS graphs start with its nodes arranged on a circle, and each vertex connected only to its $4$ closest neighbors. Then, a small fraction of these local connections is rewired to a random vertex, building a bridge far away. Nodes involved in such bridges become the most central ones because: (i) the far-away vertex will have its degree increased by $1$ and (ii) they now connect distant parts of the graph, scoring higher in betweenness, harmonic, and closeness centrality. However, these central vertices will also be close to the neighbor that has been disconnected to build the bridge. These nodes have their degree reduced by $1$, and will hence be removed earlier when looking for $k$-cores. This will in turn affect vertices nearby, including very central ones, which will hence be removed in early onion layers. In $\mathcal{G}^{WS}$, the last onion layers are occupied by nodes in lattice-like parts of the network, which the less central ones in terms of betweenness or closeness. 

      From the correlation matrix we extract Principal Components (PC) \cite{pearson1901liii}. These are directions within the space of node properties (within which the $\Pi_i$ data points live) along which nodes present more variability. In other words, this calls our attention to dimensions of our data set along which there is more heterogeneity (hence more interesting structure to report) of nodes. PC define an orthogonal basis of the space of node properties. We can project the original data into this basis---we will note such projected data as $\hat{\Pi} \equiv \{\hat{\Pi}_i,\> i=1, \dots, N^n\}$ and we will say that data is represented in PC-space or eigenspace. Note that each network node is represented as a point either in property space or PC-space. For visualization purposes, we often retain the first three PC and color-code each node according to the values they take in these components. We associate red, green, and blue to the first, second, and third PC, and interpolate linearly from hexadecimal values $00$ to $ff$ between the node scoring the least and the most along each PC. 

      \begin{figure*}[] 
        \begin{center} 
          \includegraphics[width=\linewidth]{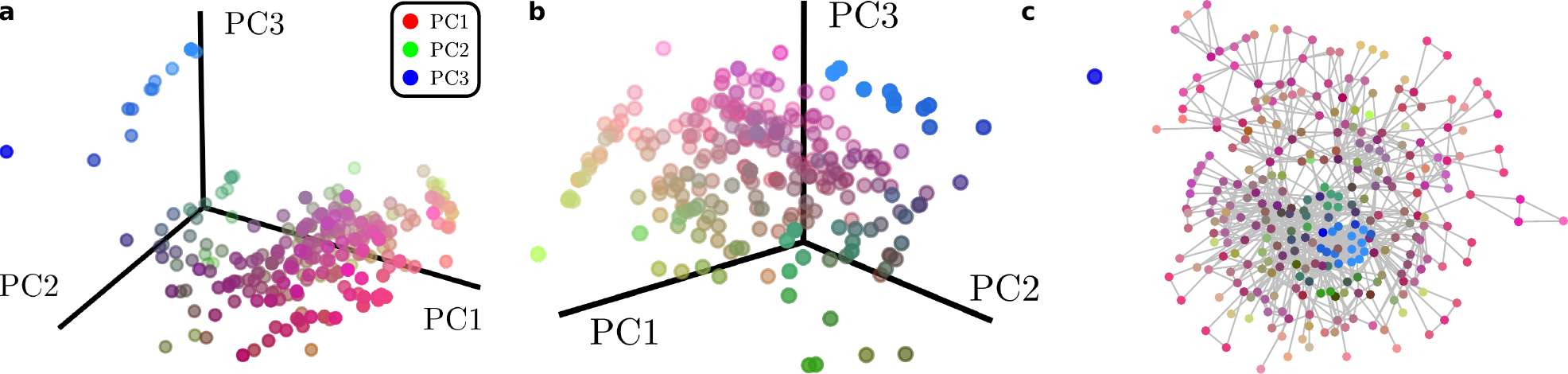}
      
          \caption{{\bf Color-coded principal components.} {\bf a}, {\bf b} Two views of the three PC for the CNB collaboration network. Each point represents one node of the network which has been color-coded according to the projection of its measured properties into the first three PC. {\bf c} PC color code projected back into the original network. }
      
          \label{fig:SM.3}
        \end{center}
      \end{figure*}

      Sup.\ Fig.\ \ref{fig:SM.3}{\bf a-b} shows this color-coded representation for the CNB collaboration network. In this space, some relevant structure becomes readily noticeable. A small set of nodes appears in blue (denoting high score in PC $3$). A valley of greener (high PC $2$) points separates the first group from the bulk of the network, which appears in redder tones (high PC $1$). Both color similarity and proximity in PC-space indicates that a group of nodes are topologically similar, meaning that they play similar structural roles in the graph. The color code can be projected back into the network (Sup.\ Fig.\ \ref{fig:SM.3}{\bf c}). This reveals that the set of bluer nodes is not only topologically similar, but it is also geometrically close. Visual inspection or classic clustering methods could have also hinted us towards this densely packed group of nodes. But, in case they miss this feature, our analysis automatically reports it because it is an outstanding one in this graph. From visual inspection or classic clustering it would have been much more difficult to find some structure in the remainder of the network, as greener and redder nodes appear rather distributed. Nodes with very similar colors can appear literally at opposite sides of the network. Despite their distance, their topological properties indicate that such vertices are deeply similar. 

      The final step of our analysis is utilizing the projection in PC-space to reveal Topological Communities (TC). This is explained in App.\ \ref{app:3}. First, let us make a brief comment on the interpretation of Principal Components. 

      Despite their widespread use and conceptual simplicity, PC might not be particularly simple to interpret. This led to the development of new methods, such as sparse PCA, that emphasize interpretability by constraining each PC to be associated with as few original properties as possible. This and other, more modern dimensionality reduction methods will advance our framework. Here we intend to introduce the TC framework with the most straightforward choices. TC study with more sophisticated techniques is left out for future work. To close this section, we illustrate some bits of information that can be extracted from PC. 

      \begin{figure*}[] 
        \begin{center} 
          \includegraphics[width=\linewidth]{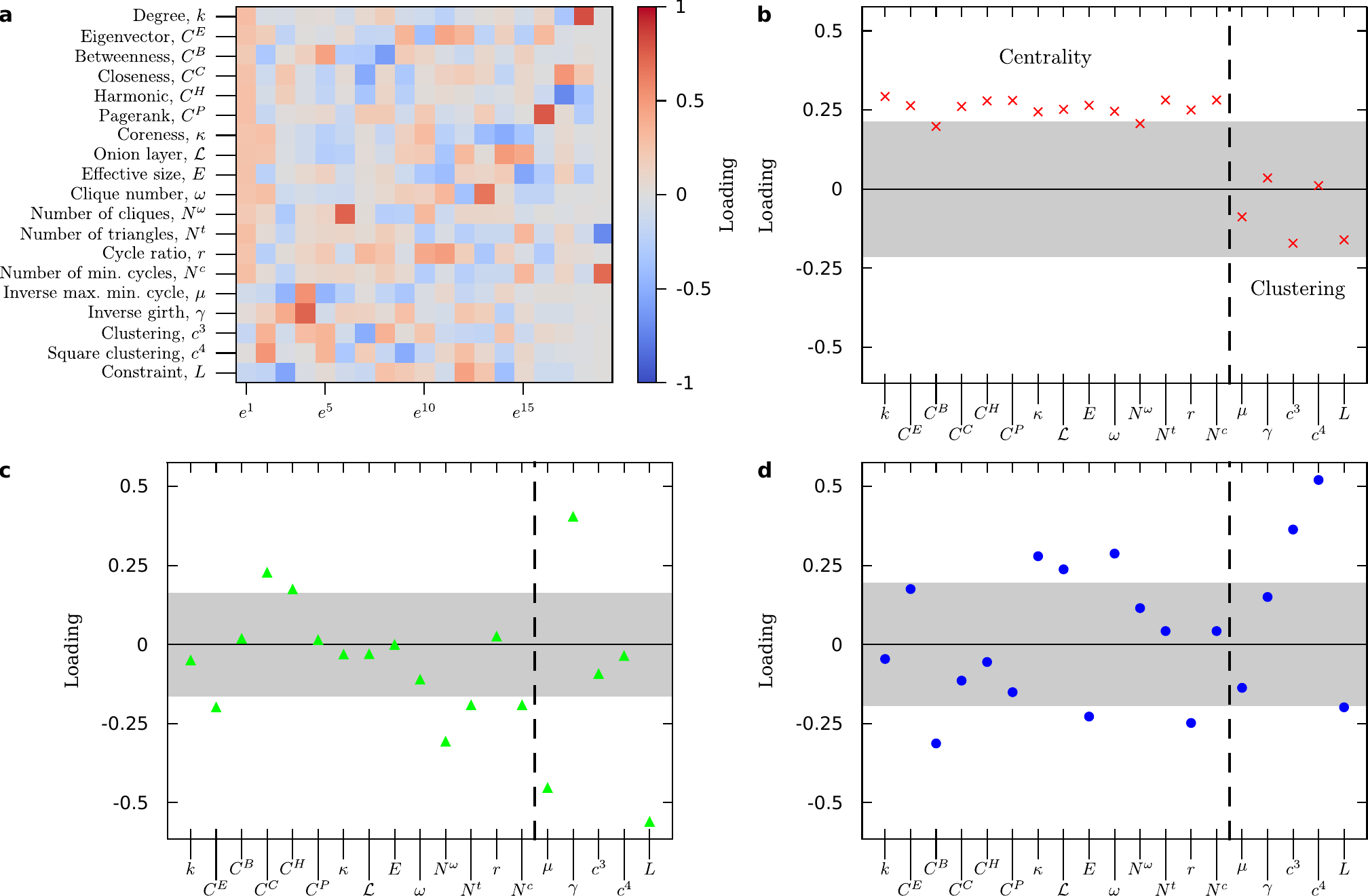}
      
          \caption{{\bf Interpreting PC of the global airport network.} {\bf a} Each column represents an eigenvector for the global airport network. Eigenvectors are sorted from left to right in decreasing eigenvalue order. Each row represents a primary property, such that the $(l, m)$ entry of this matrix is the loading of the $l$-th measured property on the $m$-th eigenvector, $\protect\overrightarrow{e}^m$. {\bf b-d} Individual plots for the first three principal components---PC are colored as in other plots. A dashed vertical line separates centrality-correlated properties from clustering-correlated properties. Shading indicates one mean of the absolute value of each eigenvector's loading above and below $0$. }
      
          \label{fig:SM.4}
        \end{center}
      \end{figure*}

      Sup.\ Fig.\ \ref{fig:SM.4} shows eigenvectors for the global airport network, $\mathcal{G}^{GTN}$. The main component (Sup.\ Fig.\ \ref{fig:SM.4}{\bf b}) correlates strongly with most properties in the centrality block from Sup.\ Fig.\ \ref{fig:SM.2}{\bf d}. This centrality defines a main axis alongside which topologically diverse nodes are segregated. This is a pattern observed in most (but not all) graphs studied, highlighting the importance of the centrality axis in complex networks. Less principal components become more difficult to decipher. The second one includes two properties related to cycles among its most relevant ones: inverse of maximum minimum cycle, $\mu$ (anti-correlated), and inverse girth, $\gamma$ (correlated). This means that nodes with both a small girth (i.e.\ a small minimum cycle associated---say a triangle) and its maximum minimum cycle is large will score higher in PC-$2$. Finally, the two main properties in PC-$3$ are the two clustering coefficients. Nodes scoring high in this component also present high core number and number of cliques, but they score low in betweenness centrality.

    \section{Locating Topological Communities}
      \label{app:3}

      We define {\em Topological Communities} (TC) as sets of nodes that are more topologically similar to each-other than to other vertices of the network. We find them by hierarchically clustering nodes that fall close in PC-space. We could use a variety of suitable techniques such as $k$-means \cite{lloyd1982least}, or methods that identify non-linear manifolds. We again opt for a straightforward method to focus on the conceptual novelty of TC. 

      We use the location of each node in PC-space to compute Euclidean distances between all vertices, then use these distances to build a dendrogram (again using fcluster from SciPy). From a bottom up perspective, this algorithm proceeds as follows: First, each node makes up its own cluster, and is represented by the node's position in PC-space. We merge the two closest nodes together in a new cluster, which becomes represented by the center of mass of the vertices just grouped up. We repeat this process iteratively, merging nodes and clusters. Nodes that are further away are merged later, inducing a distance in the emerging dendrogram. Looking at the algorithm from a top down perspective, then cutting branches at different distances along the dendrogram, a hierarchy of clusters unfolds. 

      \begin{figure*}[] 
        \begin{center} 
          \includegraphics[width=\linewidth]{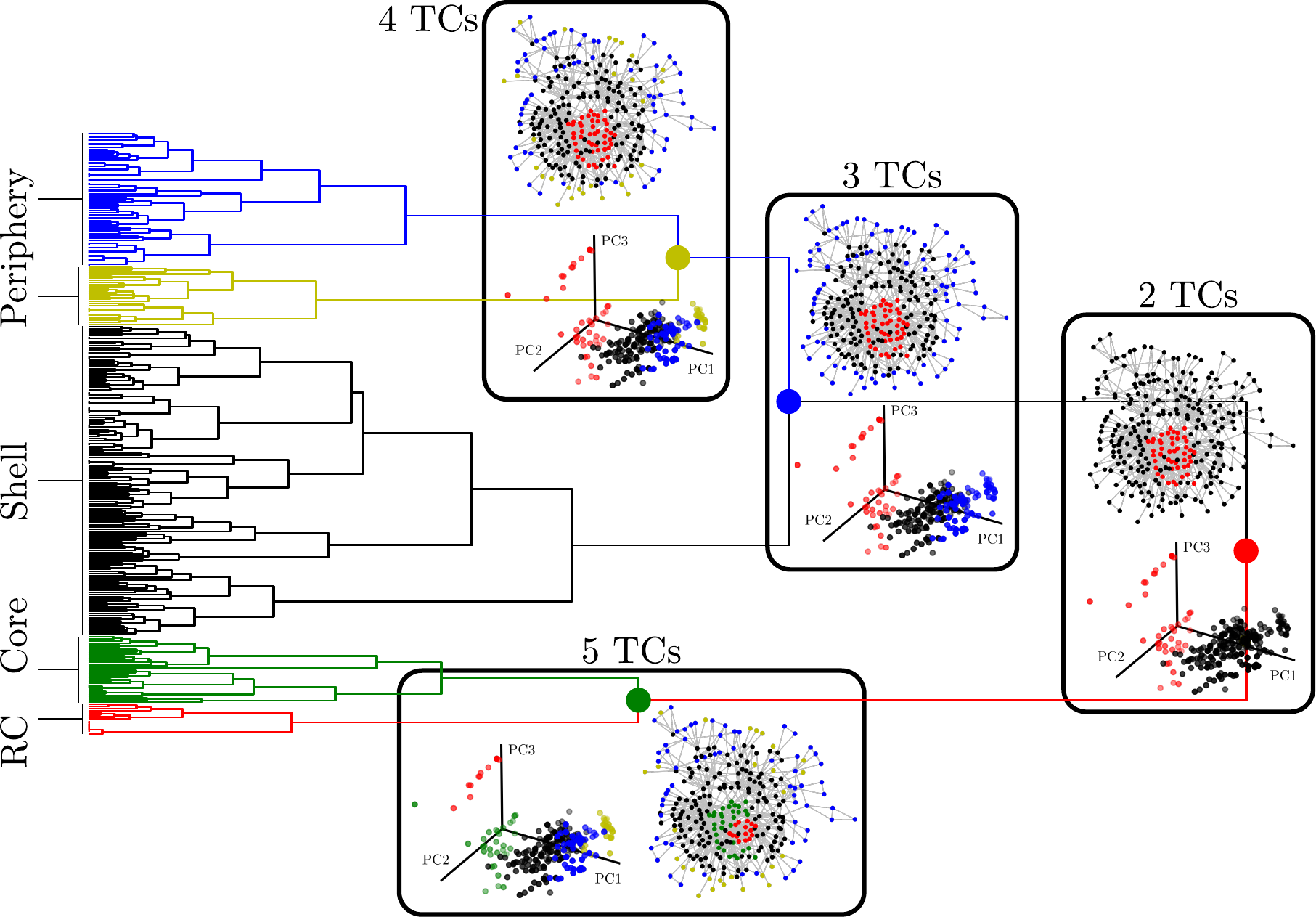}
      
          \caption{{\bf A hierarchy of TC emerges for the CNB collaboration network.} A dendrogram shows the order in which nodes are merged (left to right) by a clustering algorithm according to their distance in PC-space. In reverse (right to left) the network is split into successively smaller topological communities. We showcase the first $4$ branching events, each marked by a filled circle with the color of one of the emerging TC and accompanied by a panel showing how properties of the new TC cluster together in PC-space and how their nodes are distributed throughout the network. }
      
          \label{fig:SM.5}
        \end{center}
      \end{figure*}

      Sup.\ Fig.\ \ref{fig:SM.5} shows this process for the CNB collaboration network. Let us examine the first $4$ steps from the top down viewpoint. We start with the whole network undivided, constituting a unique cluster; then progress increasing the number of clusters as the network is split into more topologically coherent subsets of node. In Sup.\ Fig.\ \ref{fig:SM.5}, first, nodes are divided into a broad core and a periphery. Then the periphery splits twice: first revealing a shell, closer to the core; then the remainder of the nodes are separated into two kind of peripheral vertices. Finally, a subset of the core is excised revealing to constitute a rich club. 

      Let us introduce some notation before moving on. In the dendrogram, we can have any number of TC from $N^{TC}=1$ (the whole network) to $N^{TC}=N^n$ (each node constitutes its own TC). Methods to explore the optimal number of TC in each case will be explored elsewhere. In this paper we chose a suitable number in each case for illustration purposes. Let us call $\mathcal{TC}$ to the sorted set $\mathcal{TC} \equiv \{\mathcal{TC}(n),\> n=1, \dots, N^{TC}\}$ of all TC once fixed $N^{TC}$. Each TC is a collection of nodes. If a given node, $v_i$, belongs to a given TC, $TC(n)$, we will say: $v_i  \in TC(n)$ and $\mathcal{TC}(v_i) = TC(n)$. 

      \begin{figure*}[] 
        \begin{center} 
          \includegraphics[width=\linewidth]{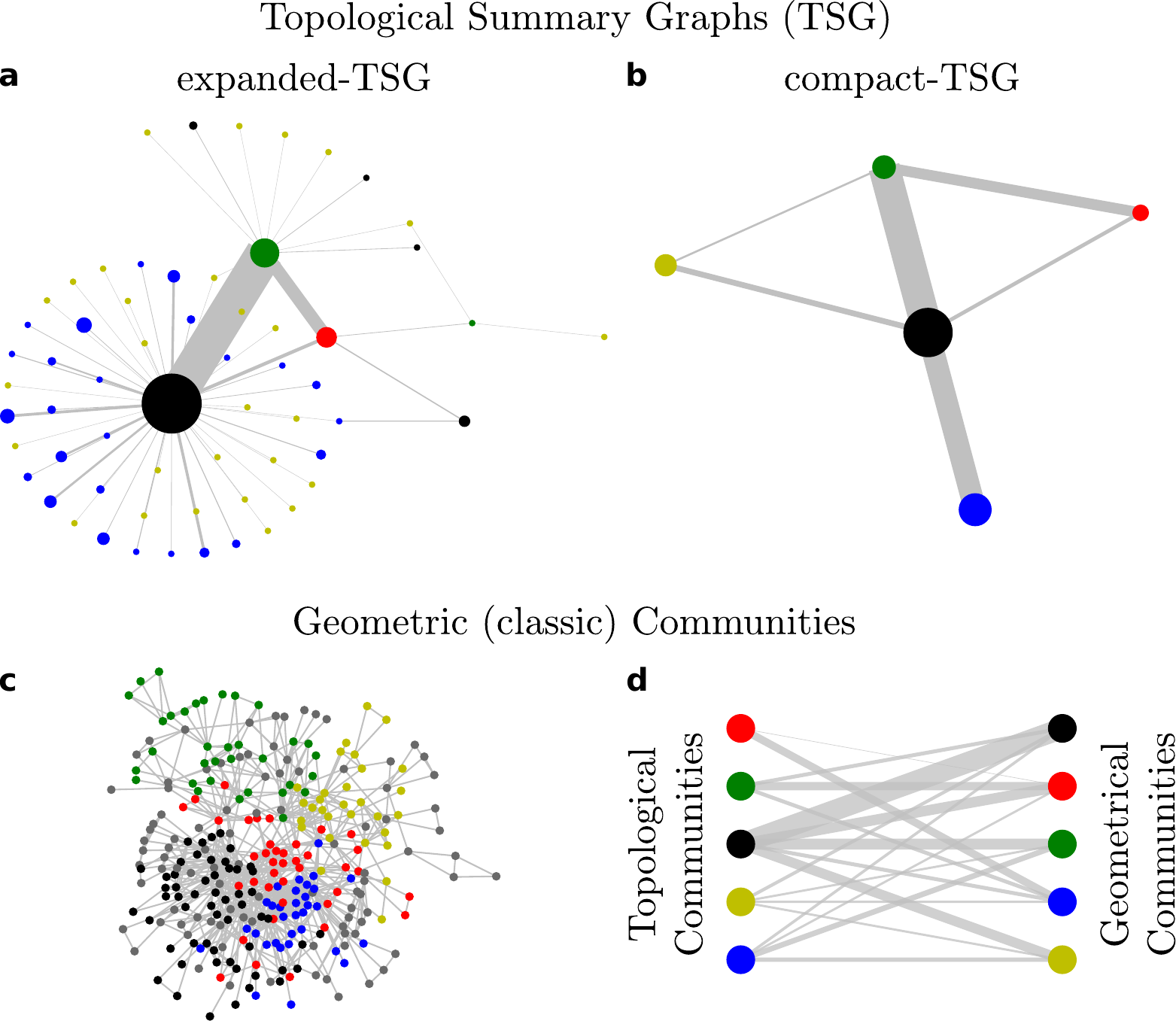}
      
          \caption{{\bf Topological summary graphs and GC for the CNB collaboration network, $\mathcal{G}^{CNB}$.} {\bf a} Expanded topologically summary graphs easily depict non-contiguous TC. Each colored node represents a subset of nodes from the main graph that cannot visit other vertices of the same TC without passing through some other TC. {\bf b} Compact topologically summary graphs contain one node for each TC and connections proportional to the number of edges across TC. {\bf c} First $5$ GC from applying a greedy algorithm to $\mathcal{G}^{CNB}$. {\bf d} Bipartite network showing how many vertices of each TC were classified into each GC, showing that topological and geometric communities extract different meaningful information about the graph. }
      
          \label{fig:SM.6}
        \end{center}
      \end{figure*}

      With this in hand, we elaborate a series of strategies to study TC. First we introduce two kinds of {\em Topological Summary Graphs} (TSG): expanded and compacted summary graphs, which capture overall relationships between TC. For expanded-TSG (eTSG), we locate all sets of contiguous nodes that belong to a same $TC$. This is, starting with an arbitrary node, locate all neighbors that belong to the same TC and that are accessible without visiting vertices that belong to another TC; etc. Mathematically, one such a set of nodes satisfies $\nu \equiv \{v_i,\> \mathcal{TC}(v_i) \land \text{ if } v_j\in V_i \text{ and } \mathcal{TC}(v_j)=\mathcal{TC}(v_i) \text{ then } v_j\in \nu\}$. Let $TC(n)$ be the TC to which all nodes in $\nu$ belong. As with individual nodes, we say $\nu \subset TC(N)$ and $\mathcal{TC}(\nu) = TC(n)$. We call $\nu$ a {\em contiguous subset} of TC(n) and we say of nodes in $\nu$ that they are {\em contiguously connected}. We make each contiguous subset into the vertices of a new graph, the eTSG, as represented in Sup.\ Fig.\ \ref{fig:SM.6}{\bf a} (elaborated for $\mathcal{G}^{CNB}$ split into $5$ TC). The size of each eTSG vertex is proportional to the number of nodes from the original network that it contains, and the width of the links is proportional to the number of edges in the original network connecting across contiguous node sets. Note that, by definition, TC impose a coloring of the eTSG, which cannot contain two adjacent vertices belonging to a same TC. 

      For compact-TSG (cTSG) graphs we group all nodes of a same TC into a same vertex disregarding their contiguity. Thus, cTSG (Sup.\ Fig.\ \ref{fig:SM.6}{\bf b} for $\mathcal{G}^{CNB}$ with $5$ TC) trivially tells us the number of nodes in each TC (vertex size) and the amount of edges connecting between different TC anywhere in the network (link width). 

      TSG for the CNB collaboration networks tell us how the most central TC (the rich club, red, and the core, green) are virtually shielded from the periphery of the nodes by the shell (black). Also, the rich club is almost completely shielded from the shell. Looking at the eTSG, we can count two green and five black vertices. This implies that neither the core nor the shell are fully contiguous sets of nodes within the network. This means that no community detection algorithm would have been able to identify the complete sets of nodes that share topological characteristics in each case, even though both these TC are large, rather central, and salient features of this graph. That task becomes even more difficult for peripheral nodes, which in the eTSG appear much more fragmented. 

      To further illustrate how TC encompass nodes that are not necessarily close, we searched for classic communities in each network. We refer to them as {\em Geometric Communities} (GC) to contrapose the defining criteria---geometric adjacency in GC versus topological similarity in TC. Sup.\ Fig.\ \ref{fig:SM.6}{\bf c} shows GC for the CNB collaboration graph found using a greedy (descriptive) algorithm. (We are aware of the troubles of such methods \cite{peixoto2021descriptive}, but, once again, we wanted to focus on illustrating TC, not technical details of GC.) Sup.\ Fig.\ \ref{fig:SM.6}{\bf d} shows a bipartite network connecting TC and GC with edge width proportional to the number of shared nodes. We have only plotted the $5$ largest GC (alongside $5$ TC) for convenience. GC algorithms group nodes based on some notion of geometric proximity within the graph (actually, we know that the three main GC slightly follow departmental divisions within the research institution from which this collaboration network is derived \cite{manrubia2022report}). They cut through TC because classic algorithms cannot separate nodes according to the topological role they play in the network.

    \section{Extended analysis of the global airport network}
      \label{app:4}

      %
      %
      We built the global transportation network, $\mathcal{G}^{GTN}$, containing the top $500$ airports (from data at \cite{marcelino2012critical, networkResources}). For each node we measured each of the primary and derived properties, finding that none is pathological (i.e.\ non yield infinity and none takes the same values for all nodes) and that we could include all of them in our analysis. We computed cross-correlation between node properties. Sup.\ Fig.\ \ref{fig:SM.2}{\bf c-d} shows the correlation matrix and dendrogram for primary properties on this network. The centrality- and clustering-correlated blocks emerge clearly. 

      \begin{figure*}[] 
        \begin{center} 
          \includegraphics[width=\linewidth]{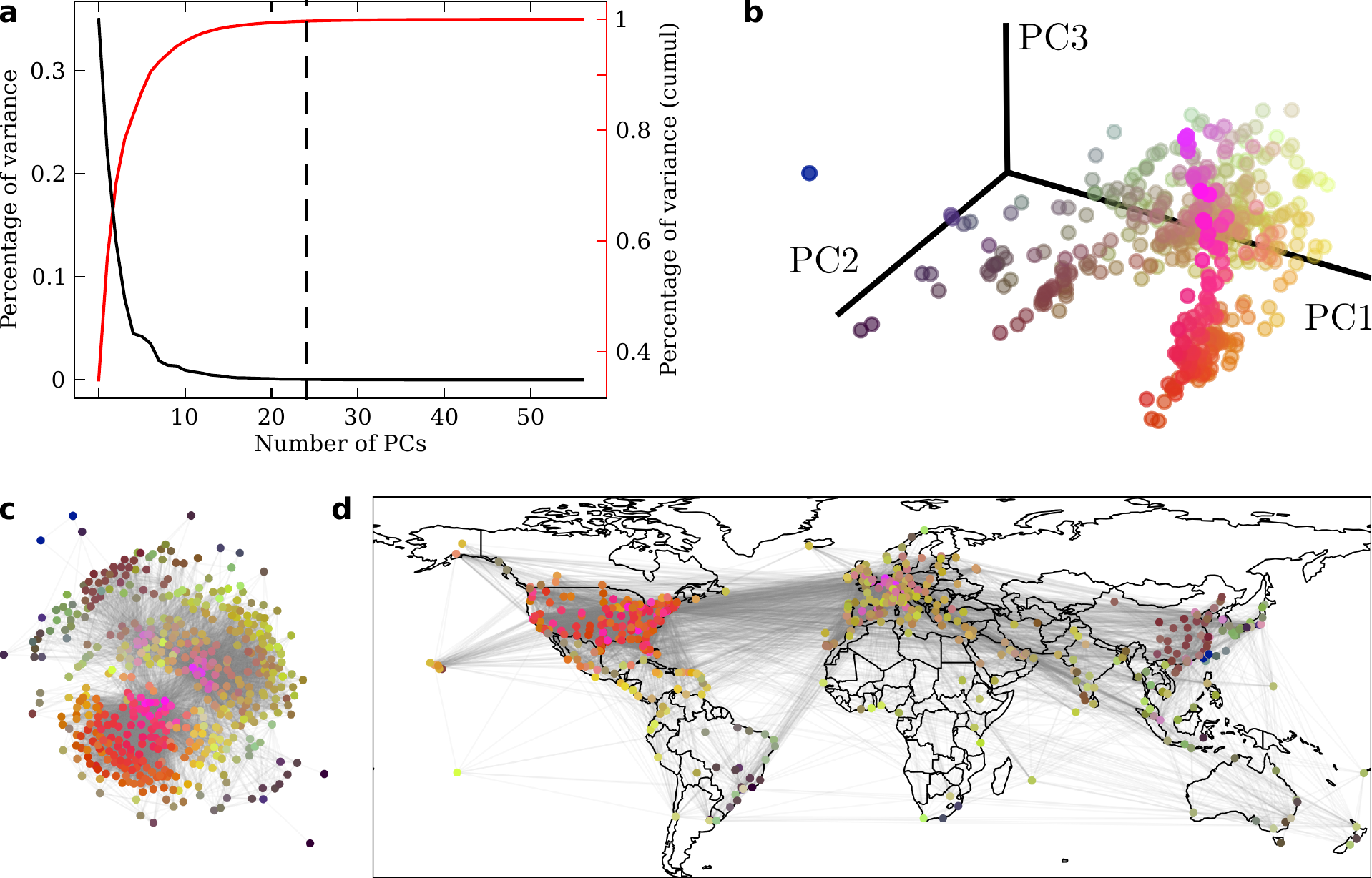}
      
          \caption{{\bf Visualizing principal components of the global transport network, . {\bf b}} {\bf a} Percentage of variance explained by each PC (black) and percentage of variance explained up to each PC (red). A vertical dashed line indicates the 24-th PC (cumulated explained variance is above $99.6\%$), after which all contribution is indistinguishable from noise (following \cite{gavish2014optimal}). {\bf b} Node properties projected into PC-space. {\bf c} PC color code projected back into network layout. {\bf d} Network visualized over the world map. Locations are those of the airports. PC color code applied. }
      
          \label{fig:SM.7}
        \end{center}
      \end{figure*}

      We diagonalize and project all node properties into PC-space (Sup.\ Fig.\ \ref{fig:SM.7}{\bf b}), where some non-trivial structure is already visible. Proximity of nodes in this space, as well as likeness in color (which codes PC $1$ to $3$ in red, green, and blue respectively), denotes similar topological characteristics within the network. When projecting this color code back into a network layout (Sup.\ Fig.\ \ref{fig:SM.7}{\bf c}) we see that some of topologically close nodes are also geometrically close---i.e.\ nearby within the network. For example, a large orange and red cluster is prominent in the center-bottom half of the network. But other vertices with similar topology are far apart in the graph---note, e.g., two brown clusters: one at the top left and a smaller one at the bottom right. Both these proximity and separation between topologically similar nodes becomes more evident when projecting the PC color code into a world map, with each node plotted at the location of the corresponding airport (Sup.\ Fig.\ \ref{fig:SM.7}{\bf d}). The reddest cluster is straightforwardly associated to the United States. Some clustering of brownish nodes appears in the South-East Asian region (around China). Nodes with a similar color can be found spread over the globe. All other colors appear extended world-wide, without an obvious clustering pattern. This anticipates that our analysis will uncover network features not based on proximity within the graph---as classic geometric communities do. 

      \begin{figure*}[] 
        \begin{center} 
          \includegraphics[width=\linewidth]{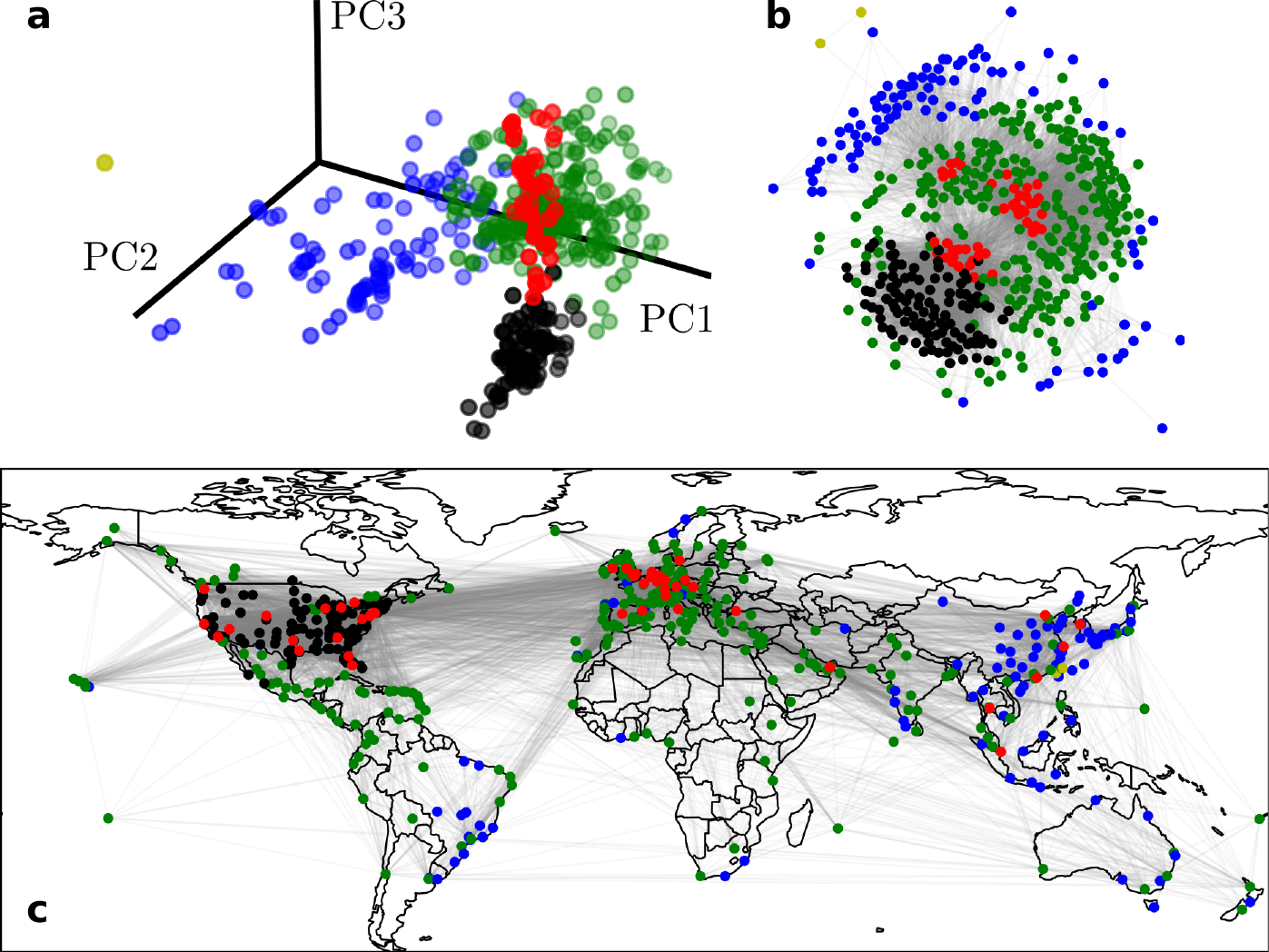}
      
          \caption{{\bf Visualizing TC of the global transport network, $\mathcal{G}^{GTN}$.} {\bf a} TC on PC-space, where they cluster together according to proximity (note that this proximity includes also less principal PC not depicted here). {\bf b} TC projected back onto network layout. {\bf c} TC projected back onto the world map reveals some interesting properties: the US-TC (black), the global backbone (red), TC-$3$ (green), and TC-$4$ (blue). }
      
          \label{fig:SM.8}
        \end{center}
      \end{figure*}

      \begin{figure*}[] 
        \begin{center} 
          \includegraphics[width=0.8\textwidth]{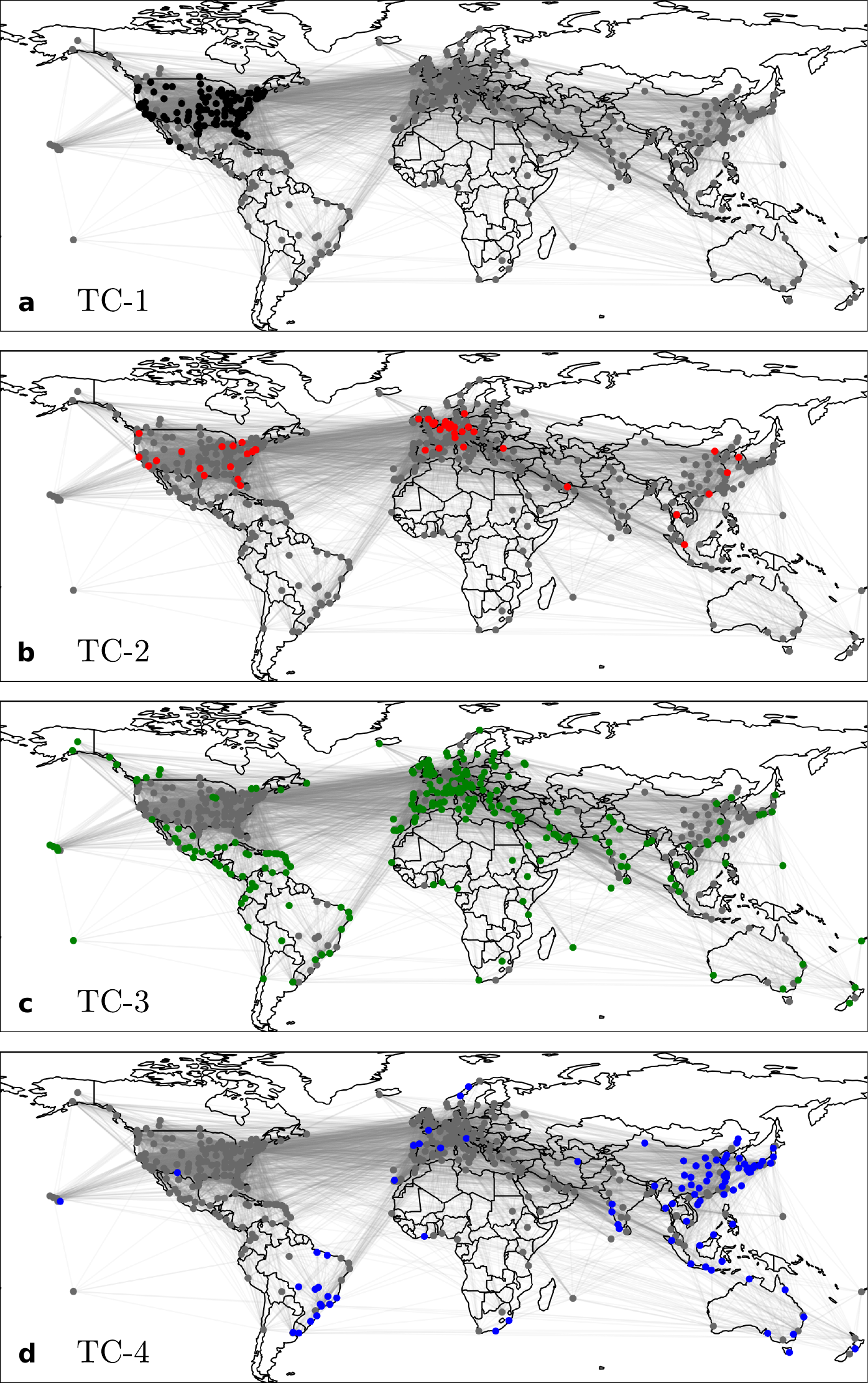}
      
          \caption{{\bf Visualizing each individual TC alone on the world map for $\mathcal{G}^{GTN}$.} {\bf a} US-TC (black) is confined to North-America. It contains few nodes in Mexico and Canada. {\bf b} The global backbone (red) keeps the network well connected. It includes the main hubs world-wide, including from the US. {\bf c} TC-$3$ (green) is the most numerous one, it is contiguous, spreads across the world, and contains most medium sized airports. {\bf d} TC-$3$ is non-contiguous. It is still well integrated in the network, which has no branches and leaves (thus not a proper periphery); but this TC is the least central one. }
      
          \label{fig:SM.9}
        \end{center}
      \end{figure*}

      \begin{figure*}[] 
        \begin{center} 
          \includegraphics[width=0.8\textwidth]{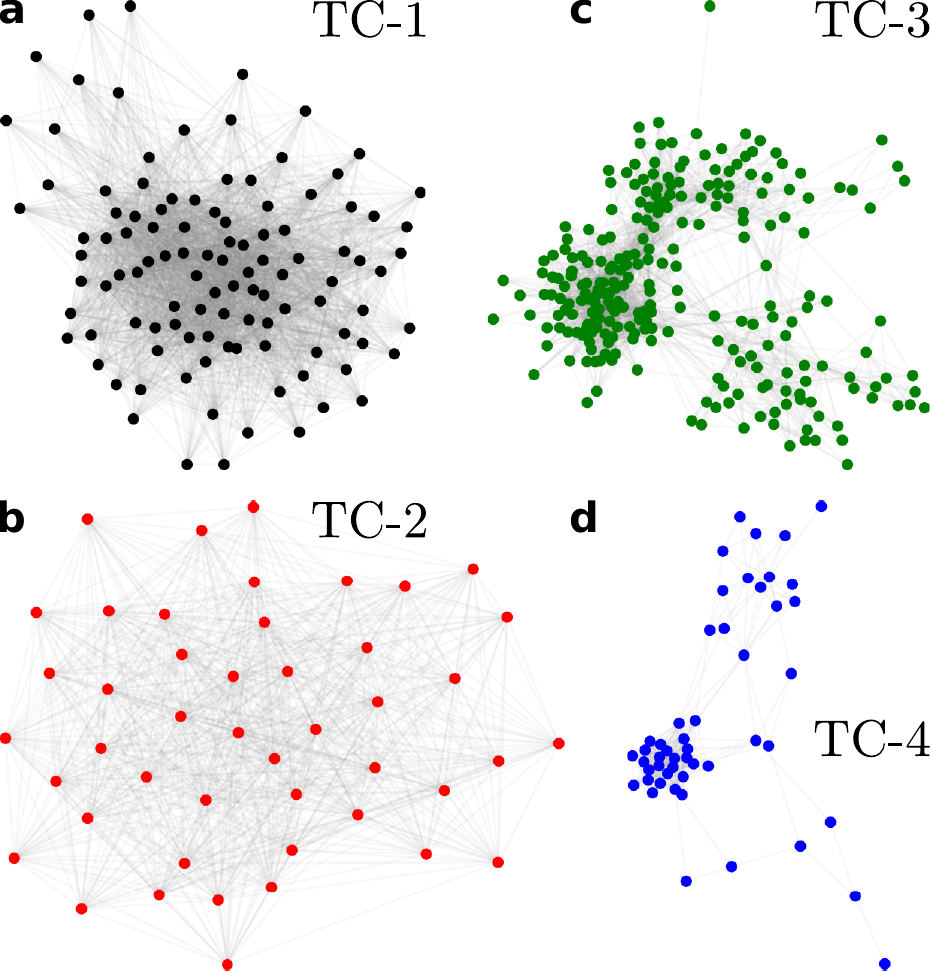}
      
          \caption{{\bf Visualizing individual TC of $\mathcal{G}^{GTN}$ as networks.} We extracted the largest connected component from each TC. Plotting them reveals different topological qualities. {\bf a} US-TC (black) seems to have some heterogeneity within, while the global backbone ({\bf b}, red) seems to be more homogeneous. {\bf c} The largest connected component of TC-$3$ seems to present some variety in structure and clusters as well. {\bf d} The largest connected component of TC-$4$ also seems to contain visible clusters. }
      
          \label{fig:SM.10}
        \end{center}
      \end{figure*}

      TC are defined by topological proximity between nodes in PC-space (Sup.\ Fig.\ \ref{fig:SM.8}{\bf a}). When projected onto a network layout, it becomes apparent that nodes in a same TC are not necessarily connected (Sup.\ Fig.\ \ref{fig:SM.8}{\bf b}). This becomes even more evident when projecting TC onto the world map (Sup.\ Fig.\ \ref{fig:SM.8}{\bf c}), and when we project each individual TC alone both on a map (Sup.\ Fig.\ \ref{fig:SM.9}) and on a graph layout (Sup.\ Fig.\ \ref{fig:SM.10}). 

      \begin{figure*}[] 
        \begin{center} 
          \includegraphics[width=0.67\linewidth]{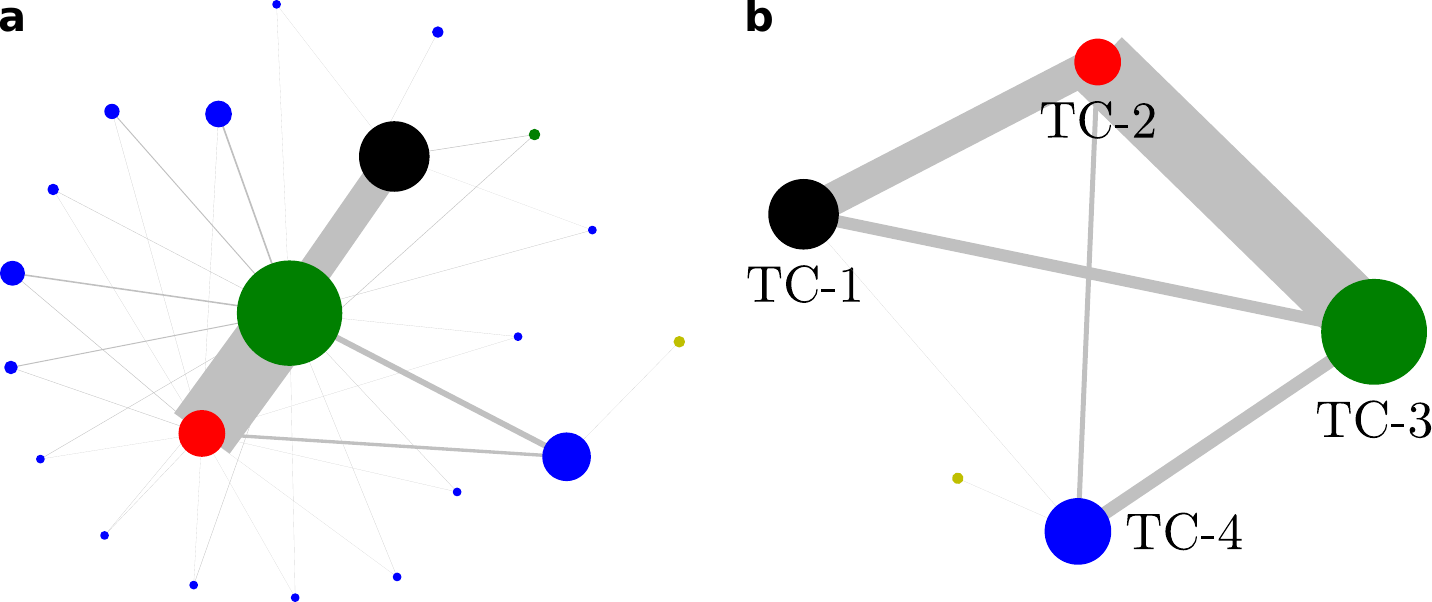}
      
          \caption{{\bf Topological summary graphs for $\mathcal{G}^{GTN}$.} {\bf a} The expanded topological summary graph reveals that both the US-TC and the global backbone are contiguous. TC-$3$ is almost contiguous (save for $2$ airports within the US) and TC-$4$ is very far from contiguous. {\bf b} The compact topologically summary graph reveals a hierarchy with the global backbone on top connecting the other two main TC. }
      
          \label{fig:SM.11}
        \end{center}
      \end{figure*}

      Sup.\ Fig.\ \ref{fig:SM.8} shows $5$ TC (we chose the number for convenience---as stated elsewhere, we will explore criteria for optimal number of TC in the future). We do not discuss TC-$5$ (hardly visible in Sup.\ Fig.\ \ref{fig:SM.10}), which consists of two poorly connected airports in Taiwan and are likely an outlier. Among the other TC, a prominent one (TC-$1$, black, Sup.\ Figs.\ \ref{fig:SM.9}{\bf a} and \ref{fig:SM.10}{\bf a}) is also spatially clustered around the United States. This shows that our analysis can report geometrically clustered nodes when they are a network feature that stands out topologically as well. This also means that a potential geometric community of US airports may also be topologically homogeneous. TC-$1$ (the US-TC) constitutes a unique contiguous subset, as illustrated in the e-TSG (Sup.\ Fig.\ \ref{fig:SM.11}{\bf a}). 

      TC-$2$ (red, Sup.\ Figs.\ \ref{fig:SM.9}{\bf b} and \ref{fig:SM.10}{\bf b}) also forms a contiguously connected components. This TC consists of the most important hubs is the world. Note that the most relevant US-airports belong here, not in the US-TC. We dub TC-$2$ the {\em global backbone} (GB) TC. This TC is denser in Europe but includes hubs in the Asian South-East and Dubai. Central and South America, Africa, India and most of the Middle East, and Oceania are not present in the global-TC, highlighting the disconnectedness of these regions from the backbone of the global transport network. Regarding its topological qualities, we see that it scores similarly to the US-TC in PC $1$ and $2$ (Sup.\ Fig.\ \ref{fig:SM.8}{\bf a}). PC-$1$ implies that both the UC-TC and the GB-TC have similar centrality features (see App.\ \ref{app:2} for interpretation of $\mathcal{G}^{GTN}$ PC). Both TC differ on PC-$3$, which correlated with clustering measurements---implying that the GB-TC has a higher clustering than the US-TC. 

      TC-$3$ (green, Sup.\ Figs.\ \ref{fig:SM.9}{\bf c} and \ref{fig:SM.10}{\bf c}) is the most widely distributed one across the world. Two US airports (Easter Iowa, CID, and the Quad Cities, MLI) belong to this TC, meaning that they are more topologically similar to airports elsewhere than to the US-TC. They constitute the only non-contiguous nodes of TC-$3$. The rest of that TC is contiguously connected (Sup.\ Figs.\ \ref{fig:SM.10}{\bf c}), as summarized by the large green node in the e-TSG (Sup.\ Fi.\ \ref{fig:SM.11}{\bf a}). Note how the main contiguous component of TC-$3$, as a graph, looks very different from those of the US-TC and the GB-TC (Sup.\ Fig.\ \ref{fig:SM.10}). This illustrates how our analysis is picking up different topological classes. 

      TC-$4$ (blue, Sup.\ Figs.\ \ref{fig:SM.9}{\bf d} and \ref{fig:SM.10}{\bf d}) looks even more dissimilar to TC-$1$ and $2$. Sup.\ Fig.\ \ref{fig:SM.10}{\bf d} shows only the largest contiguous component of this TC, which encompasses most airports in the Asian South-East. Apart from this large subset, TC-$3$ consists of $16$ contiguous components (visible in the eTDG graph, Sup.\ Fig.\ \ref{fig:SM.11}{\bf a}) spread all over the world. This TC is densely present in South America, South-East Asia, and Oceania. We remark that all nodes in TC-$4$ are topologically similar to each other despite being far away both geographically (in the world map) and geometrically (within the network). TC-$4$ scores the lowest in PC-$1$, which correlates with centrality. While the global transport network is well connected and has no terminal branches and leaves (as in a tree graph), and while it can be circumnavigated similarly to a toroid or a WS graph, if we would like to define a periphery, TC-$4$ is the best candidate. 

      The compact summary graph (cTSG, Sup.\ Fig.\ \ref{fig:SM.11}{\bf b}) shows the pattern of connections between TC, which suggests a hierarchy. On the top sits the GB-TC, that connects profusely with the US-TC and TC-$3$. These were, respectively, the topologically homogeneous transport network within the US and the most abundant and widely distributed network across the world. A few direct connections exist between the US-TC and TC-$3$, but not as many as between the GB-TC and TC-$3$. In other words, the global backbone, formed by the main hub airports, is responsible for most connections between topologically dissimilar regions of the network. This is so even though the global backbone is the smallest TC of all four. TC-$4$ is relatively disconnected from the other topological communities, and it is rather accessed from TC-$3$. 

      \begin{figure*}[] 
        \begin{center} 
          \includegraphics[width=\linewidth]{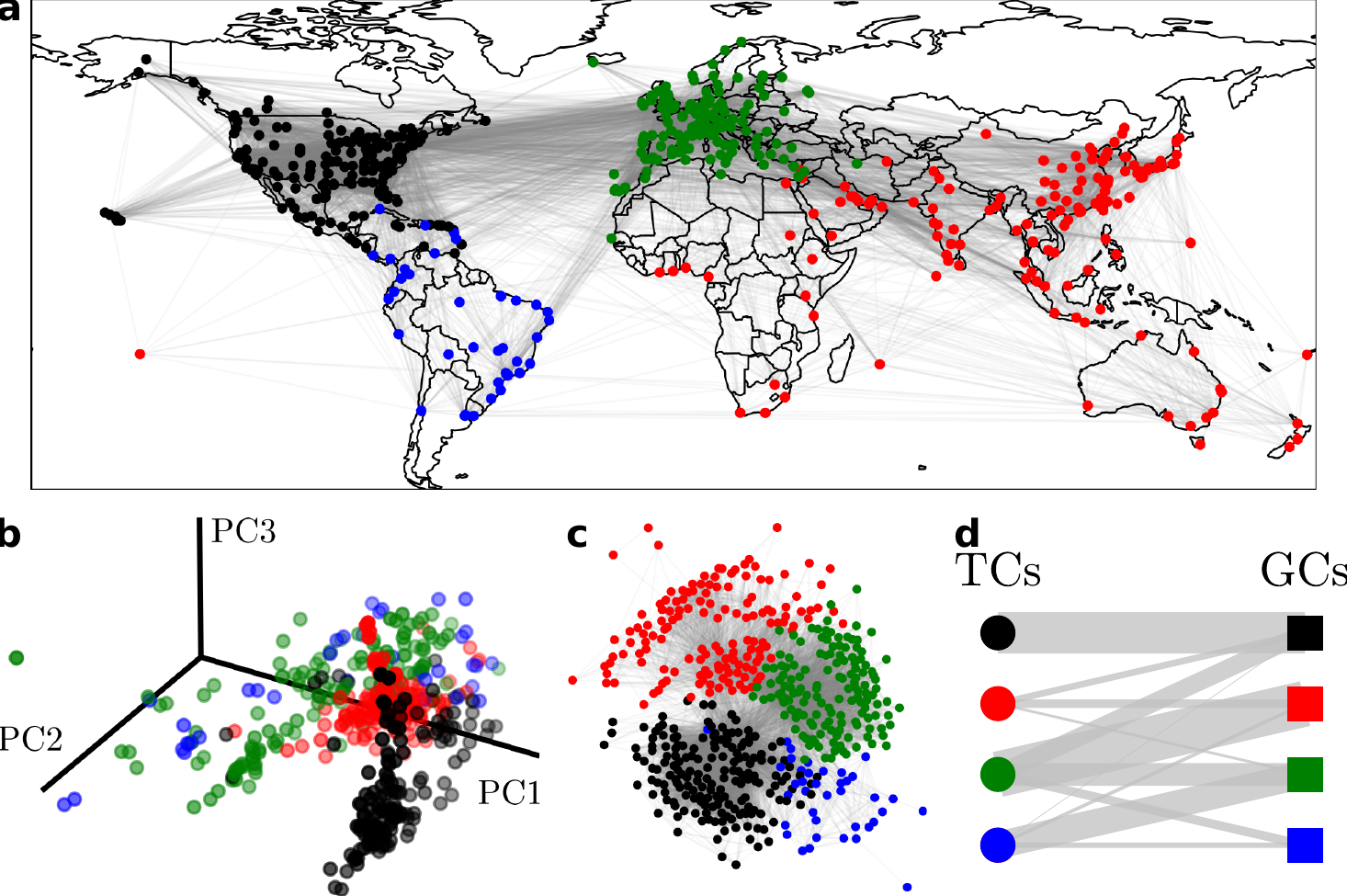}
      
          \caption{{\bf Geometric communities of the global transport network, $\mathcal{G}^{GTN}$.} {\bf a} GC are straightforward identified as geographical clusters in the world map. This geographic proximity is trivially inherited by the graph. {\bf b} GC plotted in PC-space. {\bf c} GC plotted in network layout. {\bf d} Bipartite network showing how TC are split into GC and vice versa. }
      
          \label{fig:SM.12}
        \end{center}
      \end{figure*}

      This summary of the $4$ main TC in $\mathcal{G}^{GTN}$ highlights how our analysis can group up airports that are topologically similar even if they are not close in space or in the network. This becomes much more clear if we compare TC to classic Geometric Communities within $\mathcal{G}^{GTN}$. Sup.\ Fig.\ \ref{fig:SM.12}{\bf a} shows the result of applying a greedy community detection algorithm to our graph. It reveals that nodes group preferentially according to geography, as it was already known \cite{guimera2005worldwide}. This decomposition of the graph cuts across topological categories (Sup.\ Fig.\ \ref{fig:SM.12}{\bf b}) and groups together nodes that play different roles within the graph structure at large---for example, all European airports belong to a same GC, even though only a few of them belong to the GB-TC and the rest of them are split between TC $3$ and $4$. Both analyses complement each-other, as GC look much more compact on the network layout (Sup.\ Fig.\ \ref{fig:SM.2}{\bf c}). 

      Sup.\ Fig.\ \ref{fig:SM.12}{\bf d} illustrates how TC relate to GC and vice-versa. This allows us to build an index to quantify whether topologically homogeneous communities are also geometrically close, and whether geometric communities are also geometrically homogeneous. For each TC, we compute the fraction of its nodes in each GC. We use this fraction as a probability to compute $S(n)$, the entropy of $TC(n)$ as divided into GC. Similarly, we compute the fraction of nodes from each GC assigned to each TC to compute $H(m)$, the corresponding entropy of $GC(m)$. In this example we obtain $S(1)=0$ (meaning that all airports in the US-TC are geometrically close), $S(2)=1.45$, $S(3)=1.79$, and $S(4)=1.13$. This confirms that TC-$3$ is the most widely spread TC, but both the GB-TC and TC-$4$ appear also distributed in space. On the other hand, all GC are rather evenly spread across TC, as we get: $H(1)=1.34$, $H(2)=1.32$, $H(3)=0.91$, $H(4)=0.96$. This means that, while at least one TC was able to pick up relevant geographic contiguity (as illustrated by $S(1)=0$), no GC is able to pick up topological homogeneity---not even the cluster centered in North America. 

    \section{Extended analysis of human connectomes}
      \label{app:5}

      We now study three human connectomes. The data is available at \cite{humanConnectome}. Connectomes were built by \cite{kerepesi2017braingraph} from publicly available MRI data from the Human Connectome Project \cite{humanConnectomeProject}. These networks comprise $463$ voxels of brain tissue, each of them constituting a node in the resulting graphs. The raw data gives us the number of white matter fibers inferred with standard algorithms that connect each couple of brain voxels. For our analyses, two nodes are linked if at least one fiber exists connecting the two corresponding voxels. There are several alternatives to this choice. An obvious one is to study weighted connectomes. We can also choose to connect two regions if the number of fibers between them exceeds a certain threshold (e.g.\ if two regions share more fibers than the brain-wide average). We will explore these and other alternatives in future papers. Here we stick to the simplest method. This is not crucial, since our goal is to introduce TC and showcase how they contribute to different fields (here, neuroscience). 

      The original database contains $1,064$ brains. We focus on those with patient ID $101309$ ($\mathcal{G}^{HC1}$), $992774$ ($\mathcal{G}^{HC2}$), and $989987$ ($\mathcal{G}^{HC3}$) in the Human Connectome Project database. Connectome $\mathcal{G}^{HC1}$ was chosen because it summarizes very nicely common TC found across the database after visual inspection of TC in many brains. This brain appears rather symmetric across hemispheres---as do most others in the database. In symmetric connectomes, typical TC span nodes in both hemispheres. This denotes that a given node is often more similar (in topological terms) to its contralateral partner than to other nearby regions. But this is not always the case: connectomes $\mathcal{G}^{HC2}$ and $\mathcal{G}^{HC3}$ were chosen to illustrate brain asymmetries. An asymmetric node is typically more topologically similar to others nearby than to its symmetric counterpart. Again, these three connectomes are shown to exemplify how TC can help us understand complex networked systems. More thorough analyses are underway---specifically to exploit the large database and add statistical significance to the typical TC illustrated by $\mathcal{G}^{HC1}$; or finding smaller topological communities that correlate with functional structures. 

      \begin{figure*}[] 
        \begin{center} 
          \includegraphics[width=\linewidth]{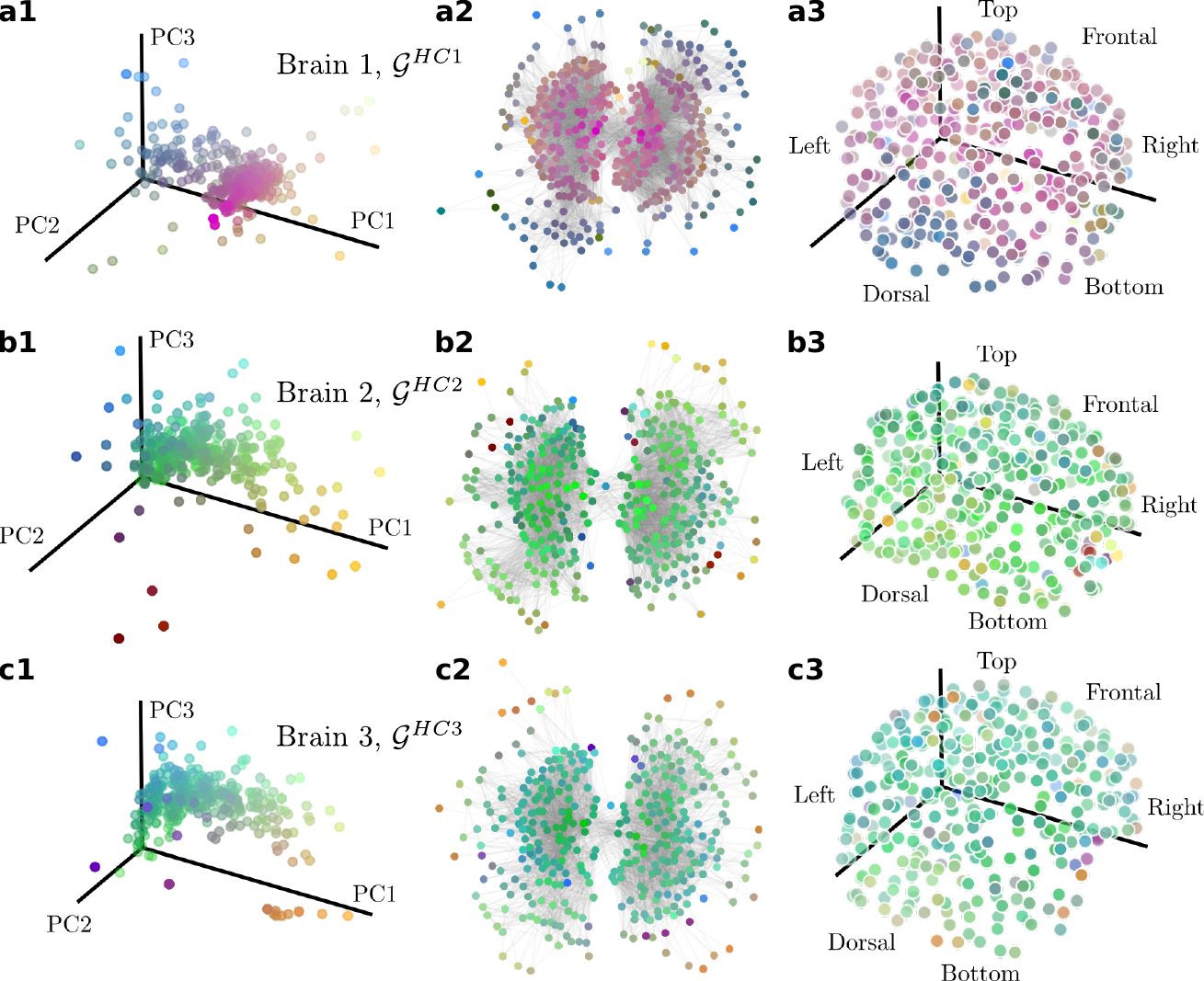}
      
          \caption{{\bf Visualizing principal components of human connectomes, $\mathcal{G}^{HC}$.} {\bf a} $\mathcal{G}^{HC1}$. {\bf b} $\mathcal{G}^{HC2}$. {\bf c} $\mathcal{G}^{HC3}$. {\bf a} Study of connectome $\mathcal{G}^{HC1}$. {\bf b} Study of connectome $\mathcal{G}^{HC2}$. {\bf c} Study of connectome $\mathcal{G}^{HC3}$. Panels {\bf 1} show distribution of each connectome's node in PC-space. Panels {\bf 2} project the PC color code back into network layout. Panels {\bf 3} project the PC color code into each node's position in the brain. }
      
          \label{fig:SM.13}
        \end{center}
      \end{figure*}

      For each connectome, we measured all primary and derived node properties finding that none is pathological (again, none takes infinite values, none takes the same value across all nodes). All were included in the analysis. Fig.\ \ref{fig:SM.13} shows the corresponding projection of nodes into PC eigenspaces (panels \ref{fig:SM.13}{\bf a1}, {\bf b1}, and {\bf c1}), and the projection of PC color code into each connectome as a network layout (panels \ref{fig:SM.13}{\bf a2}, {\bf b2}, and {\bf c2}) and on the corresponding location of each node in the brain (panels \ref{fig:SM.13}{\bf a3}, {\bf b3}, and {\bf c3}). We observe a strong contrast between colors in the first connectome (Fig.\ \ref{fig:SM.13}{\bf a}), where pink and blue hues dominate, and the second and third brains (Fig.\ \ref{fig:SM.13}{\bf b-c}), which appear mostly green. We checked that this was not due to some trivial property of PC---explicitly, whether a dimension was just sign-inverted in $\mathcal{G}^{HC1}$ with respect to $\mathcal{G}^{HC2}$ and $\mathcal{G}^{HC3}$. This was not the case. Differences come down to two reasons: (i) PC are different from one graph to another and (ii) nodes are distributed differently within each eigenspace. And this is so because each graph is different, and the TC paradigm allows each network to direct our attention to its relevant features. 

      \begin{figure*}[] 
        \begin{center} 
          \includegraphics[width=\linewidth]{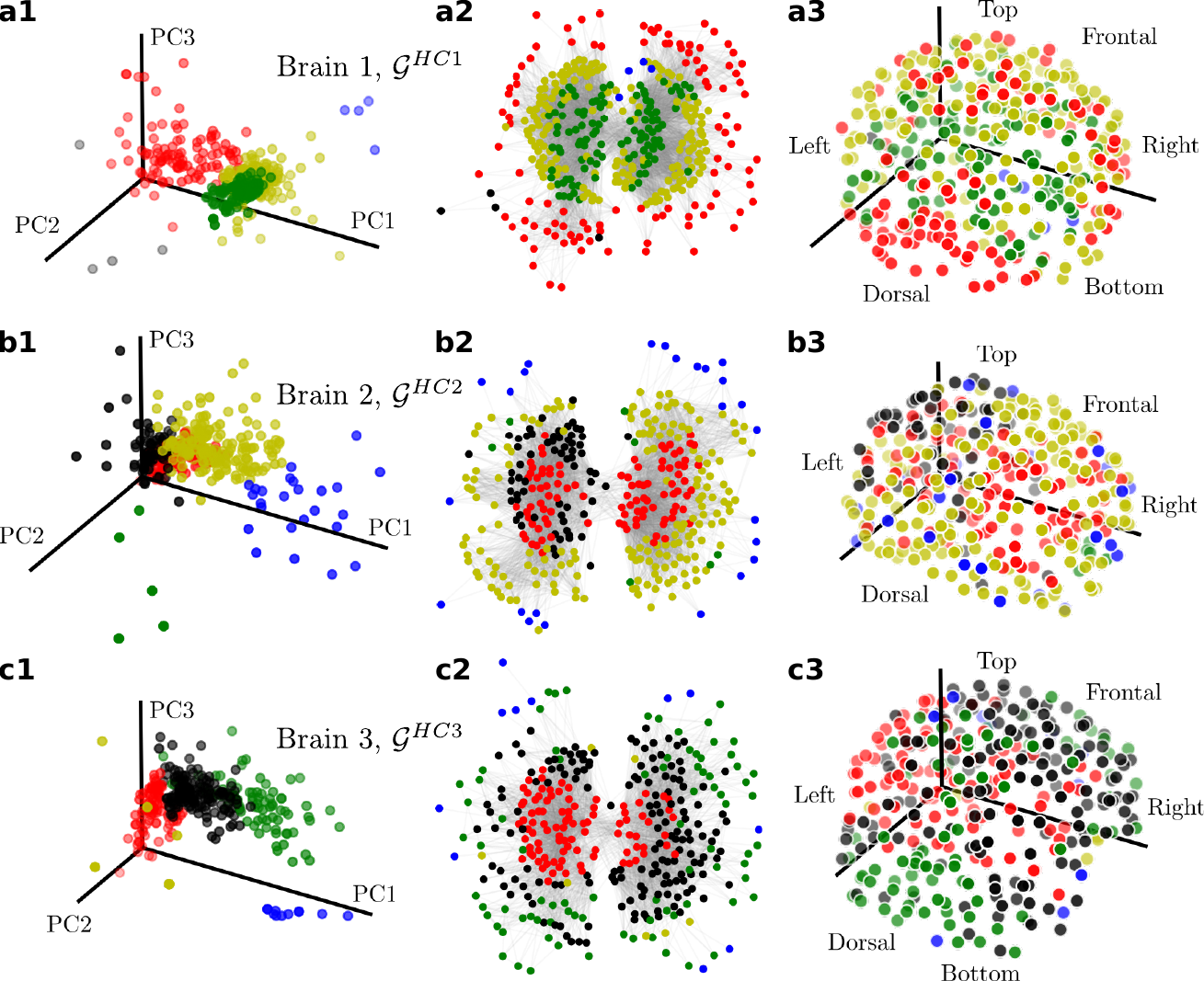}
      
          \caption{{\bf Visualizing TC of human connectomes, $\mathcal{G}^{HC}$.} {\bf a} $\mathcal{G}^{HC1}$. {\bf b} $\mathcal{G}^{HC2}$. {\bf c} $\mathcal{G}^{HC3}$. {\bf b} $\mathcal{G}^{HC2}$. {\bf c} $\mathcal{G}^{HC3}$. {\bf a} Study of connectome $\mathcal{G}^{HC1}$. {\bf b} Study of connectome $\mathcal{G}^{HC2}$. {\bf c} Study of connectome $\mathcal{G}^{HC3}$. Panels {\bf 1} show TC in PC-space. Panels {\bf 2} project TC back into network layout. Panels {\bf 3} project TC into each node's physical locations in the brain. }
      
          \label{fig:SM.14}
        \end{center}
      \end{figure*}

      \begin{figure*}[] 
        \begin{center} 
          \includegraphics[width=\linewidth]{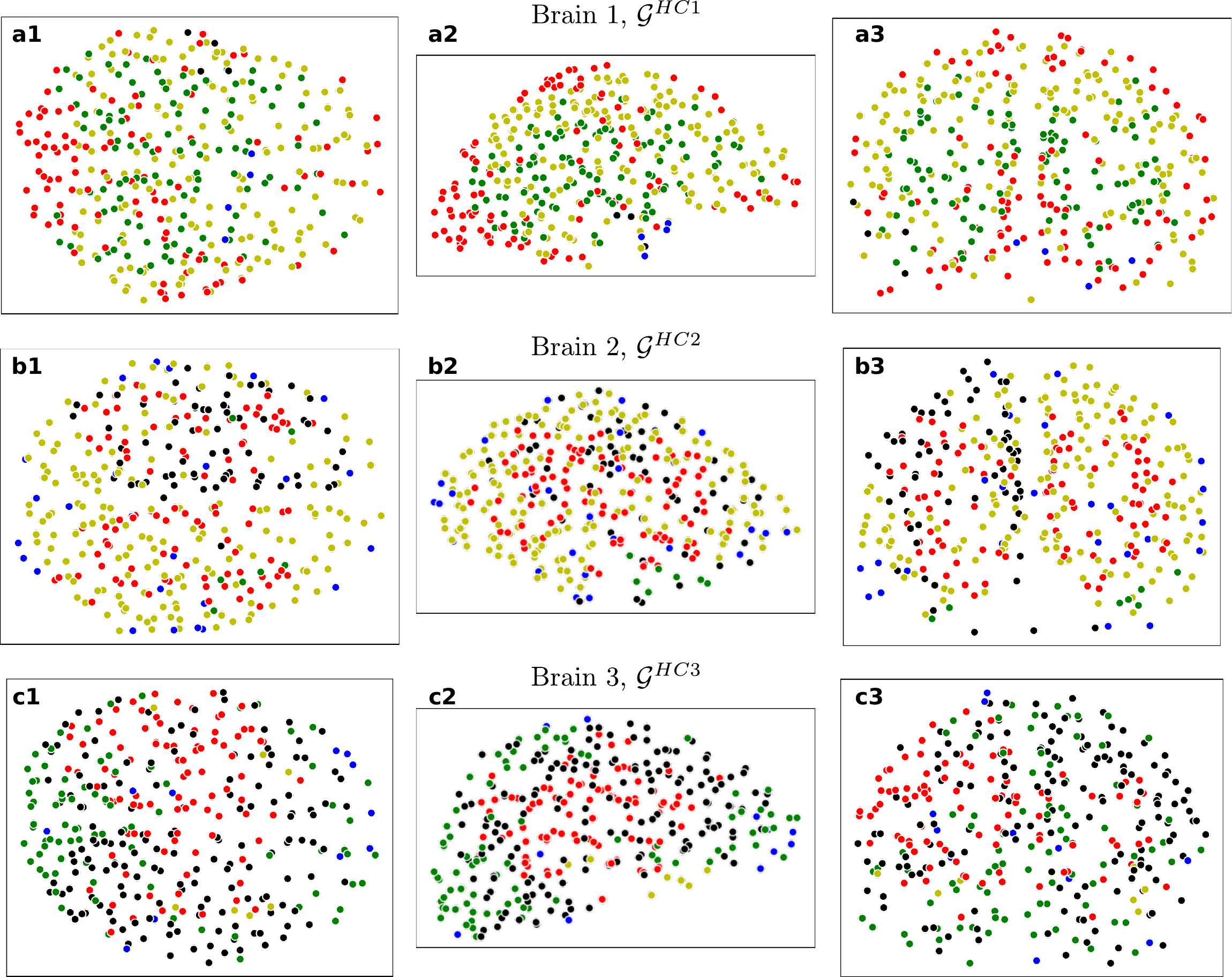}
      
          \caption{{\bf Visualizing TC of human connectomes, $\mathcal{G}^{HC}$, in the physical location of the nodes in the brain. } {\bf a} Study of $\mathcal{G}^{HC1}$. {\bf b} Study of $\mathcal{G}^{HC2}$. {\bf c} Study of $\mathcal{G}^{HC3}$. Panels {\bf 1} show coronal projection. Panel {\bf 2} show sagittal projection. Panel {\bf 3} show frontal projection. }
      
          \label{fig:SM.15}
        \end{center}
      \end{figure*}

      \begin{figure*}[] 
        \begin{center} 
          \includegraphics[width=\linewidth]{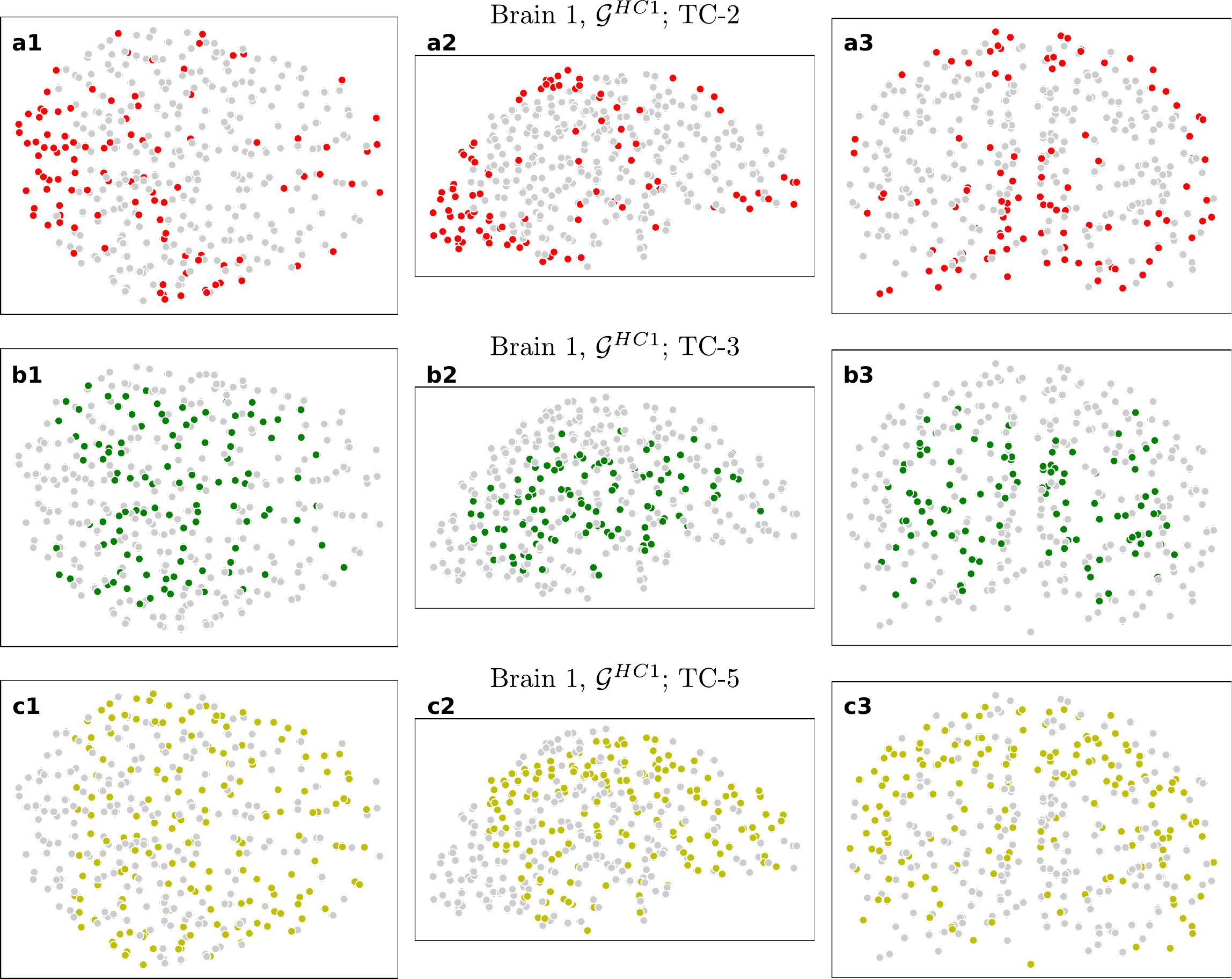}
      
          \caption{{\bf Visualizing separated TC of human connectome, $\mathcal{G}^{HC1}$. } {\bf a} TC-$2$. {\bf b} TC-$3$. {\bf c} TC-$5$. Panels {\bf 1} show coronal projection. Panel {\bf 2} show sagittal projection. Panel {\bf 3} show frontal projection. }
      
          \label{fig:SM.16}
        \end{center}
      \end{figure*}

      \begin{figure*}[] 
        \begin{center} 
          \includegraphics[width=\linewidth]{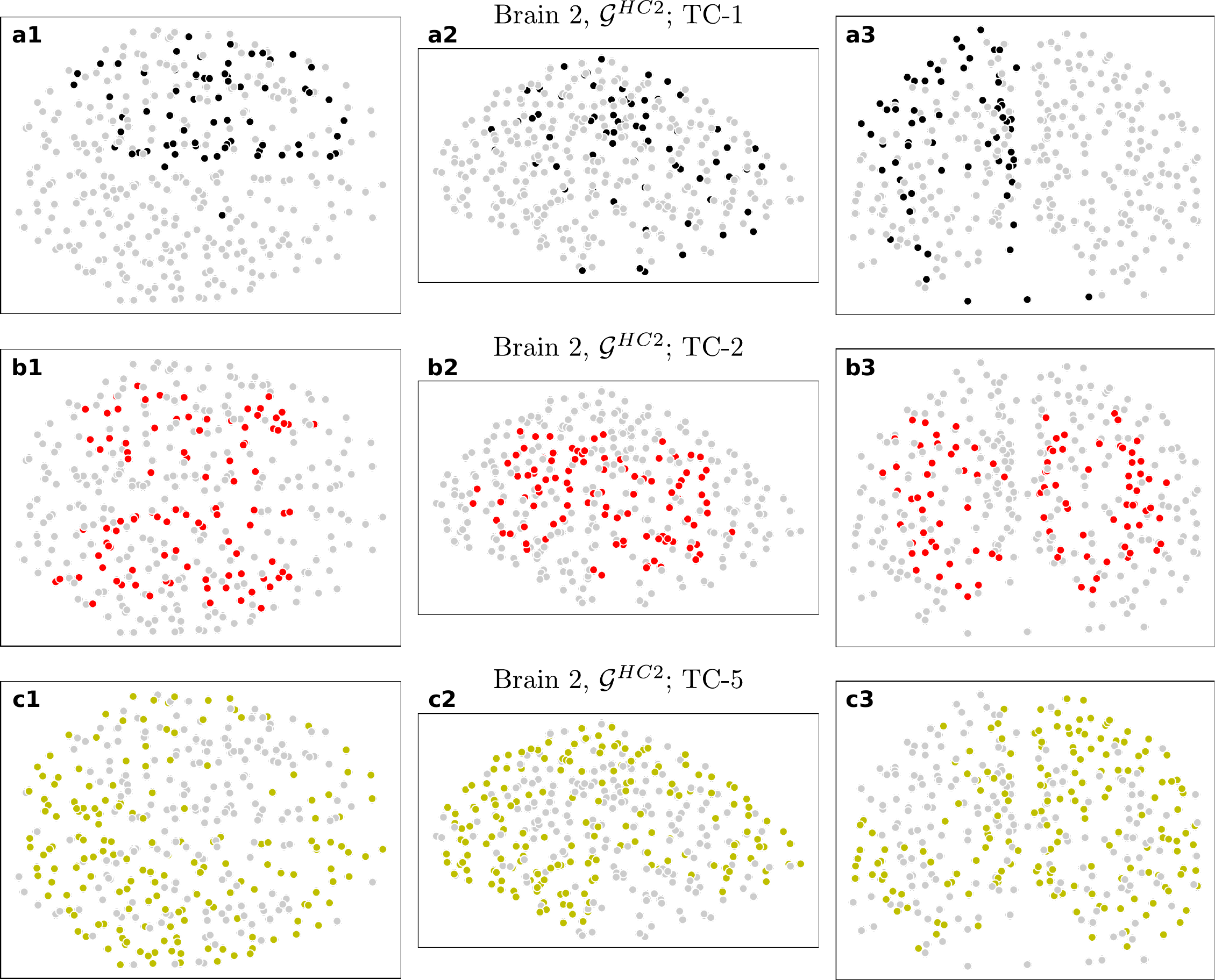}
      
          \caption{{\bf Visualizing separated TC of human connectome, $\mathcal{G}^{HC2}$. } {\bf a} TC-$1$. {\bf b} TC-$2$. {\bf c} TC-$5$. Panels {\bf 1} show coronal projection. Panel {\bf 2} show sagittal projection. Panel {\bf 3} show frontal projection. }
      
          \label{fig:SM.17}
        \end{center}
      \end{figure*}

      \begin{figure*}[] 
        \begin{center} 
          \includegraphics[width=\linewidth]{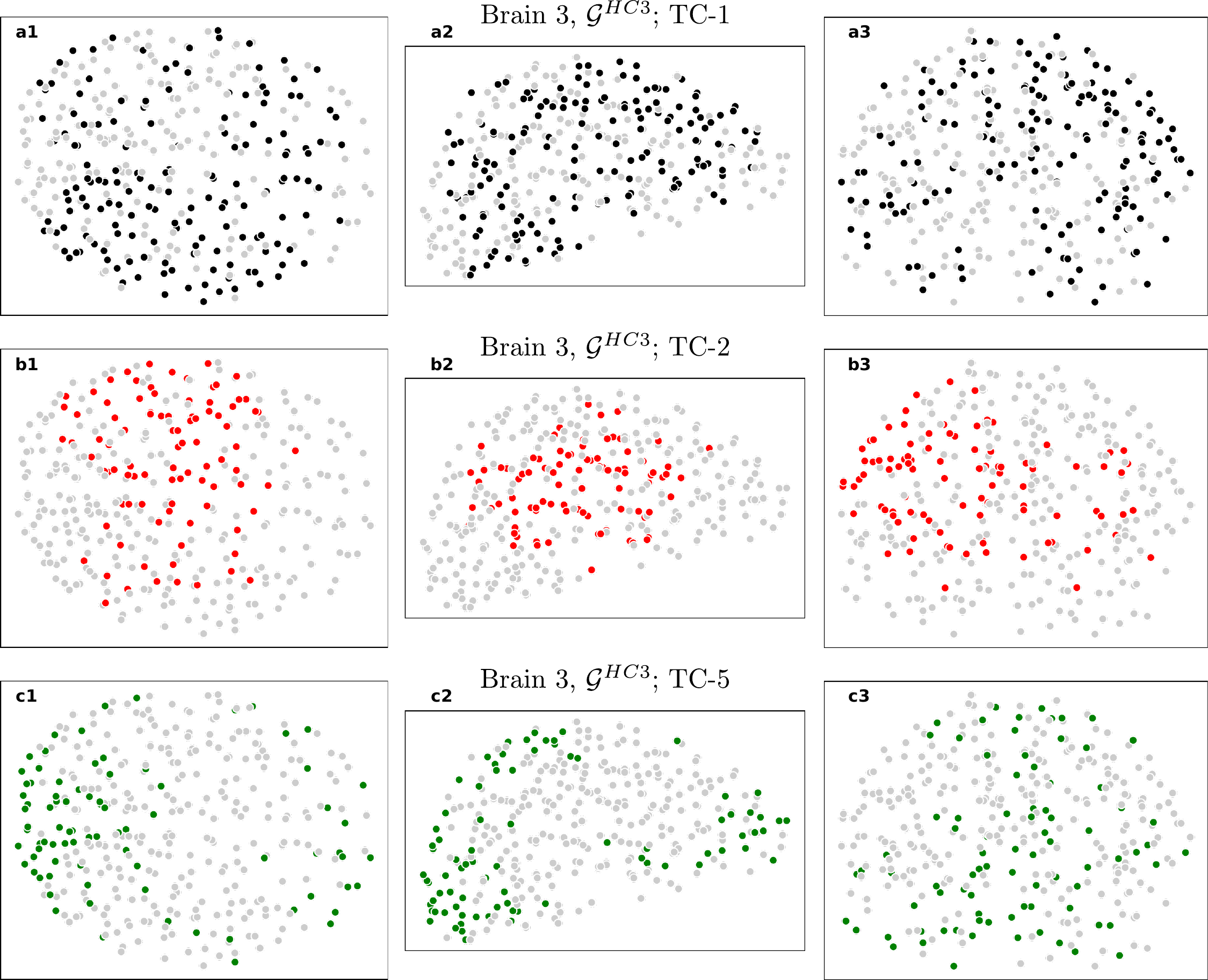}
      
          \caption{{\bf Visualizing separated TC of human connectome, $\mathcal{G}^{HC3}$. } {\bf a} TC-$1$. {\bf b} TC-$2$. {\bf c} TC-$3$. Panels {\bf 1} show coronal projection. Panel {\bf 2} show sagittal projection. Panel {\bf 3} show frontal projection. }
      
          \label{fig:SM.18}
        \end{center}
      \end{figure*}

      We advanced above that one difference concerns brain symmetry. This becomes much more clear when looking at TC, which we represent from several perspectives in Figs.\ \ref{fig:SM.14} to \ref{fig:SM.18}. Before further discussing asymmetry, let us comment on TC of $\mathcal{G}^{HC1}$, which have been observed to different degrees in most brains in the database. The most outstanding feature is TC-$2$ (red in Figs.\ \ref{fig:SM.14}{\bf a}, \ref{fig:SM.15}{\bf a}, and \ref{fig:SM.16}{\bf a}), which contains nodes mostly from the primary visual (or striate) cortex and the somatosensory area. This suggests that human connectomes around these two regions are topologically dissimilar to the rest of the brain, and that both areas are similar to each other. Next we find a set of nodes that are located in the most exposed layers of the cortex (TC-$5$, yellow in Figs.\ \ref{fig:SM.14}{\bf a}, \ref{fig:SM.15}{\bf a}, and \ref{fig:SM.16}{\bf c}). This is in opposition to another set of nodes (TC-$3$, green in Figs.\ \ref{fig:SM.14}{\bf a}, \ref{fig:SM.15}{\bf a}, and \ref{fig:SM.16}{\bf b}), that comprises mostly deeper regions. It makes sense that wiring patterns of more superficial nodes (hence remote with respect to each other) are different than those of nodes at some depth. Our analysis correctly detects this. But both TC-$3$ and TC-$5$ score remarkably similarly in the first PC (Fig.\ \ref{fig:SM.14}{\bf a1}). This is indicative of topological similarities in the wiring of most cortical nodes. In this connectome, the remaining TC-$1$ and TC-$4$ (respectively black and blue in Figs.\ \ref{fig:SM.14}{\bf a} and \ref{fig:SM.15}{\bf a}) hardly show up. They seem to correspond to deep and isolated subcortical structures, which we discuss in future work. 

      It is interesting to see how some of these structures are partly preserved in the asymmetric brains. The separation between more internal and more external regions seems present in $\mathcal{G}^{HC2}$ (Figs.\ \ref{fig:SM.14}{\bf b}, \ref{fig:SM.15}{\bf b}, and \ref{fig:SM.17}), even though the most external nodes break their symmetry into two different TC---notably captured by the corresponding TC-$1$ (black in Figs.\ \ref{fig:SM.14}{\bf b}, \ref{fig:SM.15}{\bf b}, and \ref{fig:SM.17}{\bf a}). The visual and somatosensory TC is not particularly differentiated in this brain. 

      $\mathcal{G}^{HC3}$ presents an interplay between symmetric and superficial nodes (Figs.\ \ref{fig:SM.14}{\bf c}, \ref{fig:SM.15}{\bf c}, and \ref{fig:SM.18}). Its TC-$2$ (red in Figs.\ \ref{fig:SM.14}{\bf c}, \ref{fig:SM.15}{\bf c}, and \ref{fig:SM.18}{\bf b}) contains more external nodes in the left hemisphere and more internal ones at the right. All other most external nodes are split into TC-$1$ (black in Figs.\ \ref{fig:SM.14}{\bf c}, \ref{fig:SM.15}{\bf c}, and \ref{fig:SM.18}{\bf a}) and TC-$3$ (green in Figs.\ \ref{fig:SM.14}{\bf c}, \ref{fig:SM.15}{\bf c}, and \ref{fig:SM.18}{\bf c}), which recovers the striate cortex, some of the somatosensory nodes, and other, more frontal ones. 

\begin{figure*}[] 
        \begin{center} 
          \includegraphics[width=\linewidth]{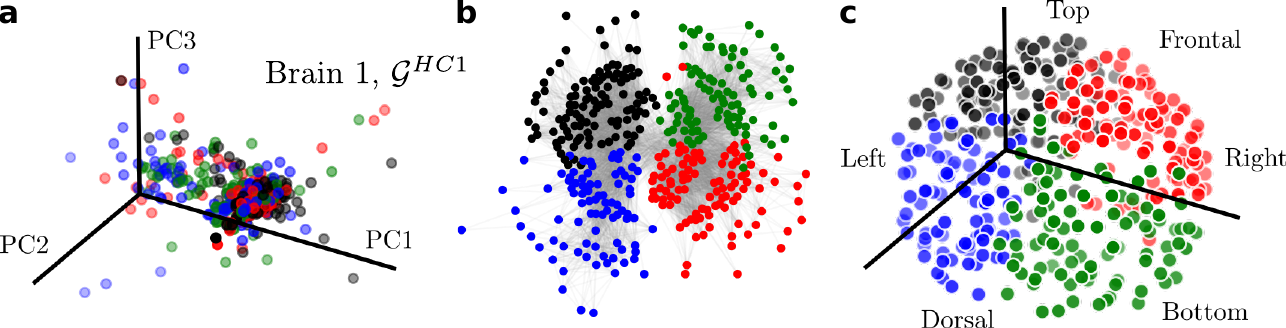}
      
          \caption{{\bf Visualizing GC of human connectome $\mathcal{G}^{HC1}$. } {\bf a} While GC on the global transport network retained some topological information, this is completely gone for connectomes. GC in PC-space mix up nodes that are nar and far in this space. But their projection in network layout ({\bf b}) and physical brain space ({\bf c}) shows good geometric coherence of GC---as expected, since that is their defining criterion. }
      
          \label{fig:SM.19}
        \end{center}
      \end{figure*}

      Looking at geometric communities in connectomes further illustrates how TC and GC are implementing two different informative decompositions of the same network. Fig.\ \ref{fig:SM.19} shows how GC in human connectomes are strongly influenced by brain geometry. Connections across hemispheres are rare in these connectomes; thus the separation across brain sides, the longitudinal fissure, constitutes a natural barrier for classical community detection algorithms. This is in stark contrast with TC. Note, e.g., TC-$3$ and TC-$5$ in the first brain, $\mathcal{G}^{HC1}$ (respectively green and yellow in Figs.\ \ref{fig:SM.14}{\bf a}, \ref{fig:SM.15}{\bf a}, \ref{fig:SM.16}{\bf b-c}). These TC group up nodes from across hemispheres. Even though connections along the longitudinal fissure are few, nodes alongside it have connectivity patterns and general topological properties similar to those of either TC-$3$ or TC-$5$, and are consequently grouped therein. In turn, GC group nodes according to their hemisphere first (Fig.\ \ref{fig:SM.19}{\bf c}). Then, within each side, nodes are split roughly into the frontal and parietal lobes, on the one hand, and the occipital and temporal lobes, on the other. 

      If we project these GC back into the eigenspace of topological properties, we see that GC does not capture any of the information structure in this space (Fig.\ \ref{fig:SM.19}{\bf a}). This indicates that, for human connectomes, classical community detection algorithms are missing out on a large share of meaningful information about these graphs. Note that this was not completely so for the global transport networks, where one of the TC presented large overlap with a classical community. In general, we cannot assume that this will happen, and both methodologies should be pursued to obtain complementary information about each network. 

    \section{Brief analyses of additional networks}
      \label{app:6}

      We include some extra case studies to illustrate the diversity of topological decompositions that can be found. These analyses are briefer than the previous ones---we will expand some of them in dedicated papers. The range of topologically distinct networks obtained comes from considering just unweighted and undirected graphs, even though some of them are naturally weighted, directed, or both. We expect to uncover more topological diversity and deeper insights when including this information in future analyses. This is beyond the scope of this paper, which intends to introduce and illustrate the core concept of TC. \\

      \begin{figure*}[] 
        \begin{center} 
          \includegraphics[width=\linewidth]{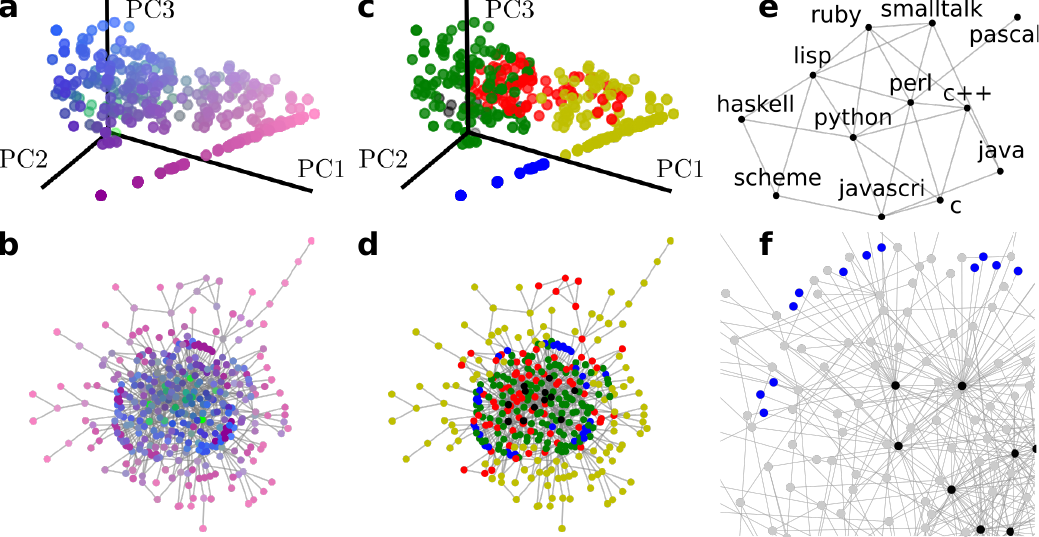}
      
          \caption{{\bf A brief study of TC in a programming languages network, $\mathcal{G}^{PL}$.} {\bf a} Nodes define a manifold in PC-space that looks different from those of earlier graphs. {\bf b} PC-space color code projected back into network layout. {\bf c} TC as defined by proximity in TC space. {\bf d} TC projected back into network layout help us clarify their roles. {\bf e} A backbone (sparser than that of the global transport network) sits at the center of $\mathcal{G}^{PL}$. It contains the most successful programming languages. {\bf f} TC-$4$ (blue) contains a periphery of nodes that have no descendants but are directly connected to the backbone. }
      
          \label{fig:SM.20}
        \end{center}
      \end{figure*}

      Fig.\ \ref{fig:SM.20} shows a TC analysis for programming languages. The network was elaborated in \cite{valverde2015punctuated} based on whether a programming language is based on another---e.g.\ C++ is trivially based on C, but less obvious relationships were spelt out in \cite{valverde2015punctuated}. This network is unweighted but directed, which we ignore. 

      %
      %
      From the projection of nodes into PC eigenspace (Fig.\ \ref{fig:SM.20}{\bf a}) we find a cloud of points, or manifold, different to all previous ones. This indicates that the topological structure of this graph is different from other examples. At the very center of the network we find TC-$1$ (black), which we denominate the graph's backbone. This TC contains very few languages, hence it is hardly noticeable in PC eigenspace (black nodes hidden in the back ground between green nodes in Fig.\ \ref{fig:SM.20}{\bf c}), but its central place is apparent in network layout (Fig.\ \ref{fig:SM.20}{\bf d}). It contains the most relevant programing languages (Fig.\ \ref{fig:SM.20}{\bf e}), from which all other structured languages descend. Note a relevant difference between this backbone and the one in the global transport network, $\mathcal{G}^{GTN}$: nodes within the $\mathcal{G}^{GTN}$ backbone were very tightly connected, almost completing a clique (Fig.\ \ref{fig:SM.10}{\bf b}); but the ones in Fig.\ \ref{fig:SM.20}{\bf e} is more sparsely connected, yet it holds the graph together. 

      Another interesting aspect of this network is that it presents two peripheries: a large one, TC-$5$ (yellow in Fig.\ \ref{fig:SM.20}{\bf c-d}), and a smaller one, TC-$3$ (blue). This last TC is singular (and different from the two peripheral TC found for the CNB collaboration network) in that all nodes descent exclusively from backbone languages. They are either very recent or very unsuccessful nodes that have not inspired newer programming languages yet. The TC analysis manages to pick up this interesting topological feature. A classical community detection algorithm would likely group these nodes somewhere alongside backbone languages, despite their deep differences. \\

      \begin{figure*}[] 
        \begin{center} 
          \includegraphics[width=\linewidth]{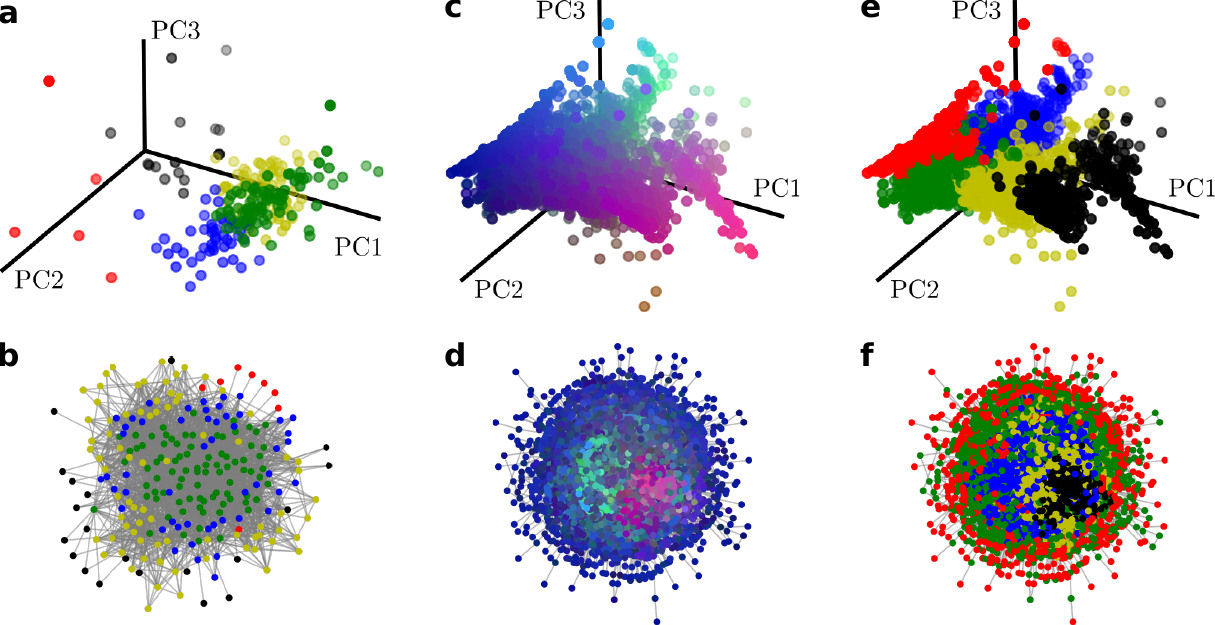}
      
          \caption{{\bf Brief TC studies of a Macaque connectome, $\mathcal{G}^{MC}$, and a yeast protein-protein interaction network, $\mathcal{G}^{Y}$.} {\bf a} TC in PC-space for the macaque connectome, $\mathcal{G}^{MC}$. {\bf b} $\mathcal{G}^{MC}$ TC projected back into network layout. {\bf c} Nodes of a yeast's protein-protein interaction network, the largest in our analyzed in this paper, shows a very densely populated manifold. {\bf d} PC color code projected back into the yeast network layout. Intriguing clusters arise that will be studied in the future. {\bf e} TC in PC-space for yeast protein-protein interaction. {\bf f} TC projected back into network layout. }
      
          \label{fig:SM.21}
        \end{center}
      \end{figure*}

      Fig.\ \ref{fig:SM.21} summarizes the analysis of a macaque brain connectome---built in \cite{harriger2012rich} with collated data from $410$ tract tracing studies. This way of connectome reconstruction is different from the techniques used for human connectomes. Among other things, we are presented with a composition of several macaque brains. The projection in PC eigenspace (Fig.\ \ref{fig:SM.21}{\bf a}) might bear some resemblance to that of the typical human connectome, $\mathcal{G}^{HC1}$ (Fig.\ \ref{fig:SM.13}{\bf a}). But a lot of its features are absent. Noticeably, when plotted in network layout (Fig.\ \ref{fig:SM.21}{\bf b}), it does not display the marked division between hemispheres in human brains (e.g.\ Fig.\ \ref{fig:SM.13}{\bf a2}). A possibility is that tract tracing is able to recover many more inter-hemispheric connections than MRI, thus reducing the chasm. \\ 

      The largest network that we have processed so far is the protein-protein interactome of the yeast {\em Saccharomyces cerevisiae}, $\mathcal{G}^{Y}$. This network has very recently been presented with exquisite detail \cite{michaelis2023social}. It contains $3,839$ nodes and $30,955$ edges, both an order of magnitude larger than all other networks in this study. 

      Network size is a current limitation of our analysis. On the one hand, it is time-consuming to compute all topological properties for each node. Some calculations scale quickly with network size---e.g.\ betweenness centrality grows as $\sim N_n^3$; other properties, even faster. On the other hand, complex networks often present heavy-tailed distributions---notably so for node degree. Heavy tails often appear also in measurements that correlate with degree, such as the different centralities. Thus, for very large networks, the first PC is usually dominated by heavy-tailed variables. This can eclipse more interesting topological features which distributions decay exponentially, and eventually skews TC detection too. This effect takes on a very visual form: eigenspace projections of networks with heavy-tailed properties result in a few nodes stretching the first PC by orders of magnitude,  while all others dimensions appear flattened. A possible solution is to take logarithms for heavy-tailed properties, which are more informative in these cases. This would allow other features to have the relevance they deserve in defining TC. 

      It has been possible to obtain an informative TC decomposition of $\mathcal{G}^{Y}$ because all its properties behave nicely despite being such a large graph---i.e.\ no heavy tails. Fig.\ \ref{fig:SM.21}{\bf c} shows a densely populated eigenspace. The cloud of points appears different than in other examples, again revealing a distinct topological disposition of nodes. Features in eigenspace are readily assigned to TC (Fig.\ \ref{fig:SM.21}{\bf e}), which are non-trivially distributed over the network (Fig.\ \ref{fig:SM.21}{\bf f}). The study in \cite{michaelis2023social} also provides a detailed map of proteins within the yeast cell. This will allows us to link TC to structure and function in future studies. \\ 

      Complex networks have been of great aid in understanding social systems \cite{milgram1967small, travers1977experimental, palla2005uncovering}. A recent, fruitful case study has been the US house of representatives \cite{neal2014backbone, andris2015rise, neal2020sign, hohmann2023quantifying}, in which voting members can collaborate to sponsor a same bill. We can build a graph, $\mathcal{G}^{US}$, that connects representatives who sponsored together more bills than expected by random. Such networks have helped uncover the pattern of polarization currently in full sway in the US and elsewhere in the world. Applying our TC analysis we can reproduce results concerning polarization and extract some new insights. 

      \begin{figure*}[] 
        \begin{center} 
          \includegraphics[width=\linewidth]{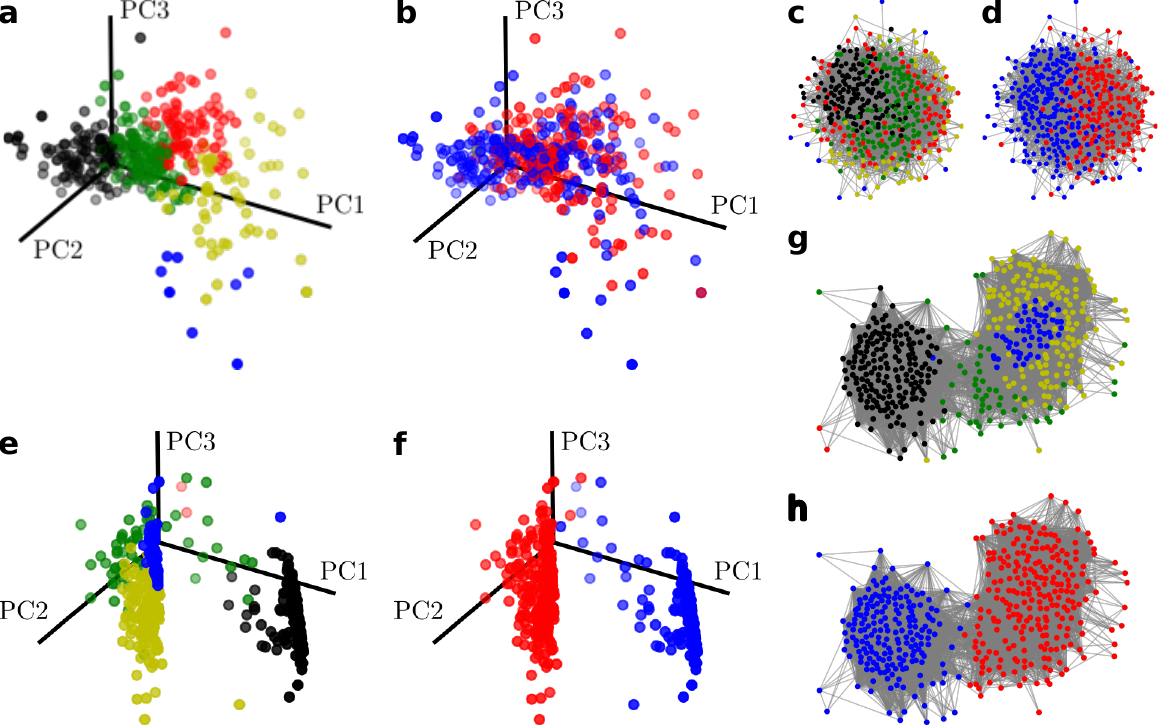}
      
          \caption{{\bf Brief study of two US house bill co-sponsorship networks, $\mathcal{G}^{US93}$ and $\mathcal{G}^{US114}$.} {\bf a-d} Study of $\mathcal{G}^{US93}$. {\bf e-h} Study of $\mathcal{G}^{US114}$. Panels {\bf a} and {\bf e} show TC in PC-space, where very different topology already differentiates the two networks. Panels {\bf b} and {\bf f} show nodes in PC-space divide according to party lines. Panels {\bf c} and {\bf g} Show TC projected back onto network layout. Panels {\bf d} and {\bf h} show nodes colored according to party in network layout. } 
      
          \label{fig:SM.22}
        \end{center}
      \end{figure*}

      Fig.\ \ref{fig:SM.22}{\bf a-d} explores the bill co-sponsorship network in the US house of representatives for the 93rd congress (Jan 1973 to Jan 1975), $\mathcal{G}^{US93}$. Fig.\ \ref{fig:SM.22}{\bf e-h} shows the same analysis for the 114th congress (Jan 2013 to Jan 2015), $\mathcal{G}^{US114}$; forty years apart. The evolution toward the current, polarized state is remarkable and it leaves a clear imprint both in PC eigenspace and in TC. Note how division after party lines in $\mathcal{G}^{US93}$ (Fig.\ \ref{fig:SM.22}{\bf b}) does not result in a clear topological separation of nodes. This means that, within both parties, there are numerous representatives that took on similar topological roles within this social network. Opposed to this, in $\mathcal{G}^{US114}$, dividing nodes along party lines results also in a meaningful segregation of topological properties (Fig.\ \ref{fig:SM.22}{\bf f}). In other words, representatives of a party occupy different topological roles within $\mathcal{G}^{US114}$---so much so that this chasm is revealed by the naked eye in PC eigenspace. 

      In future studies, we intend to apply the TC paradigm to data from the intervening years. But an interesting insight is revealed with the current, limited analysis. In the 114th congress, with well advanced social polarization, and with salient topological clusters associated to either party, there is a symmetry breaking between the two largest groups. The Democrats (blue in Fig.\ \ref{fig:SM.22}{\bf f}, {\bf h}) is made up of mostly one large TC (TC-$1$, black in Fig.\ \ref{fig:SM.22}{\bf e}, {\bf g}). This indicates that most bill-support patterns established by Democrats in the 114th congress are very similar to each other. Meanwhile, Republicans (red in Fig.\ \ref{fig:SM.22}{\bf f}, {\bf h}) can be decomposed into three prominent TC: (i) TC-$3$ (green in Fig.\ \ref{fig:SM.22}{\bf e}, {\bf g}), which contains representatives ready to collaborate across party lines (including a few Democrats); (ii) TC-$4$ (blue in Fig.\ \ref{fig:SM.22}{\bf e}, {\bf g}), which contains representatives very central to the Republican subnetwork; and (iii) TC-$5$ (yellow in Fig.\ \ref{fig:SM.22}{\bf e}, {\bf g}), which contains what seems a large periphery of Republican representatives. 

      Note that TC in this example summarize bill-support patterns from both parties. Since more republicans appear in the bridge TC, we may wonder whether Republicans may be more eclectic and ready to collaborate than Democrats, as this symmetry breaking of the TC decomposition suggests. This seems not to be the case. Analysis of the 111th congress ($\mathcal{G}^{US111}$, not shown) portrays a reversal between Democratic and Republicans topological decompositions---Republican have less topological diversity than Democrats in $\mathcal{G}^{US111}$; and vice-versa, with Democrats splitting into $3$ TC that include a bipartisan collaboration subnetwork. We hypothesize that the difference stems from who has the majority (Democrats in the 111th congress; Republicans across the 112th-114th congresses, in which topological patterns are as in Fig.\ \ref{fig:SM.22}{\bf e-h}). This suggests that the minority party adopts a more homogeneous strategy for legislative collaborations, while the majority party may be forced to have representatives playing different roles. Clarifying this becomes relevant only in the current, polarized scenario. These questions are less important in the 93rd congress because representatives of either party are more topologically similar to each other. These and other issues will be investigated in successive papers.

    \section{Connection with earlier studies}
      \label{app:7}

      Our work is directly inspired by methods currently in vogue in neuroscience \cite{gallego2017neural, houston2022squishy, gardner2022toroidal, sebastian2023topological} or cell and molecular biology \cite{becht2019dimensionality}. These recent contributions have explored complex neural or biological systems by studying large collections of objects (e.g.\ activity of neural assemblies, electric brain waves, or different cells), each of these objects being described by an also large collection of qualitative features (e.g.\ neurons involved in the assembly activity; frequency, intensity and other dynamical markers of neural waves; arrays of gene expression in cells). Those objects of interest become hence represented by points in very high-dimensional spaces. But features often correlate with each other, thus the seemingly complex, high-dimensional cloud of points made up by all objects in a study can often be summarized by a very low dimensional manifold. This manifold can be inferred through dimensionality reduction techniques such as PCA, uMap, and others. 

      Take \cite{gardner2022toroidal} to illustrate these methods at their best. Gardner et al.\ recorded activity from a very large assembly of neurons in the entorhinal cortex of mice while they moved freely across an empty maze---i.e.\ a flat plane. It is known that these regions code the mouse's position. As a consequence, the binary vector that indicates if each neuron is ON or OFF at a point in time also moves around a high-dimensional space, its trajectory tracking the mouse's whereabouts in the abstract mathematical realm. Those binary collections constitute feature vectors. Adjacent points in physical space often produce similar representations, hence correlations arise. When applying dimensionality reduction techniques, neural assembly activity turns out to dwell in a relatively simple $2$-D surface: a toroid. This manifold happens to capture, quite optimally, relevant properties of positions over a plane. This has implications for theories of neural representation, as discussed in \cite{gardner2022toroidal}. 

      In this work we turned these trendy, powerful methods to finding complex graph decompositions that are based on explicit topological similarity between individual nodes. The resulting high-dimensional representation of each network can often be reduced to very low dimensional manifolds, each with a characteristic shape that summarizes the diverse topological roles played by vertices. Because our decompositions (TC) group nodes based on topological similarity, vertices with similar topological roles do not need to be (and often are not) contiguous in the graph. This lies beyond the capabilities of classic community detection techniques, which fail to capture the highly informative structure found by TC. 

      The need for better characterization of distinct topological roles within classic geometric communities, and the recognition that nodes with a same topological role might be present in different GC and that they might not be contiguous, has been made explicit in some papers that inspired our work. Notably, in \cite{guimera2005worldwide}, classic geometric communities are detected in a global airport transport network (akin to ours). The (now-)expected clustering of airports by geographical zones is discovered, but then an additional analysis is carried out to find distinct roles within such geographic clusters. For example, some airports within a country are hubs connecting to the wider global network, while others are part of a provincial periphery. This prior knowledge might guide our intuition suggesting we measure within-module degree or participation in outside clusters (as the authors in \cite{guimera2005worldwide} did) to separate nodes according to these more refined topological aspects. 

      We expand this kind of analysis in several ways. We want each network to tell us what is salient in its topology. We do not assume that sets of nodes will stand out because of some aspect of degree, or any other specific quantity. Instead, we try to capture all potentially relevant graph-defining facets of nodes. Dimension reduction techniques then guide us to each graph's salient structures in a less biased way. Thus, ours is a more principled and encompassing framework to address this problem in complex networks. 

      Other recent approaches have tried to group nodes also not based on classic geometric communities, but rather on similarities between their neighborhoods. This can be important, e.g., to identify plausible functional routes in genetic regulation. Even if two genes belong to different regulatory modules, if they target similar nodes downstream, they might be functionally related or perhaps be making use of a same signaling route for completely different purposes. This is a relevant aspect of network topology, and methods such as introduced in \cite{pascual2020functionink} take care of it. The TC framework should be able to extract similar information if and when such structures are relevant aspects of the studied graph. Note that for some research we might not be interested in retrieving, in order of saliency, all TC. We might be just interested, e.g., in overlap between targeted nodes (for which the framework in \cite{pascual2020functionink} is preferred). But if we care about the topological make up of a graph at large, then the TC paradigm improves by incorporating more different, relevant dimensions and by detecting less contiguous, yet similar sets of nodes (e.g.\ not necessarily connected to a same nighborhood, but sharing othere abstract similarities). 

      Finally take the case of node2vec, an elegant and popular approach to make nodes and network structure readable to neural networks. This technique starts a series of random walks from each node, recording the sequence of vertices visited. These sequences are then provided to artificial neural networks that try to extract patterns between random walks, then groups up nodes according to the patterns detected. Again, node2vec might rely on features that are shared by proximal nods---since they are more likely to produce similar random walks. On the other hand, node2vec and the neural networks it employs act much as black boxes---we do not control what patterns might be extracted from random walks, nor whether they exhaustively cover all relevant topological or geometrical dimensions of a graph. While efforts exists to make interpretable AI, we think that by focusing on specific topological aspects (which are readily interpretable and central to what makes graphs different from each other), the TC framework is a more straightforward contribution to expand our understanding of complex networks.

\vspace{0.2 cm}

	\bibliographystyle{ieeetr}

\begin{thebibliography}{10}

		\bibitem{shen2002network}
			S.~S. Shen-Orr, R.~Milo, S.~Mangan, and U.~Alon, ``Network motifs in the
			transcriptional regulation network of escherichia coli,'' {\em Nature
			genetics}, vol.~31, no.~1, pp.~64--68, 2002.

		\bibitem{farkas2003topology}
			I.~Farkas, H.~Jeong, T.~Vicsek, A.-L. Barab{\'a}si, and Z.~N. Oltvai, ``The
			topology of the transcription regulatory network in the yeast, saccharomyces
			cerevisiae,'' {\em Physica A: Statistical Mechanics and its Applications},
			vol.~318, no.~3-4, pp.~601--612, 2003.

		\bibitem{luscombe2004genomic}
			N.~M. Luscombe, M.~Madan~Babu, H.~Yu, M.~Snyder, S.~A. Teichmann, and
			M.~Gerstein, ``Genomic analysis of regulatory network dynamics reveals large
			topological changes,'' {\em Nature}, vol.~431, no.~7006, pp.~308--312, 2004.

		\bibitem{dobrin2004aggregation}
			R.~Dobrin, Q.~K. Beg, A.-L. Barab{\'a}si, and Z.~N. Oltvai, ``Aggregation of
			topological motifs in the escherichia coli transcriptional regulatory
			network,'' {\em BMC bioinformatics}, vol.~5, pp.~1--8, 2004.

		\bibitem{palla2005uncovering}
			G.~Palla, I.~Der{\'e}nyi, I.~Farkas, and T.~Vicsek, ``Uncovering the
			overlapping community structure of complex networks in nature and society,''
			{\em nature}, vol.~435, no.~7043, pp.~814--818, 2005.

		\bibitem{chen2008revealing}
			Z.~J. Chen, Y.~He, P.~Rosa-Neto, J.~Germann, and A.~C. Evans, ``Revealing
			modular architecture of human brain structural networks by using cortical
			thickness from mri,'' {\em Cerebral cortex}, vol.~18, no.~10, pp.~2374--2381,
			2008.

		\bibitem{bullmore2009complex}
			E.~Bullmore and O.~Sporns, ``Complex brain networks: graph theoretical analysis
			of structural and functional systems,'' {\em Nature reviews neuroscience},
			vol.~10, no.~3, pp.~186--198, 2009.

		\bibitem{van2012high}
			M.~P. Van Den~Heuvel, R.~S. Kahn, J.~Go{\~n}i, and O.~Sporns, ``High-cost,
			high-capacity backbone for global brain communication,'' {\em Proceedings of
			the National Academy of Sciences}, vol.~109, no.~28, pp.~11372--11377, 2012.

		\bibitem{harriger2012rich}
			L.~Harriger, M.~P. Van Den~Heuvel, and O.~Sporns, ``Rich club organization of
			macaque cerebral cortex and its role in network communication,'' 2012.

		\bibitem{betzel2014changes}
			R.~F. Betzel, L.~Byrge, Y.~He, J.~Go{\~n}i, X.-N. Zuo, and O.~Sporns, ``Changes
			in structural and functional connectivity among resting-state networks across
			the human lifespan,'' {\em Neuroimage}, vol.~102, pp.~345--357, 2014.

		\bibitem{montoya2006ecological}
			J.~M. Montoya, S.~L. Pimm, and R.~V. Sol{\'e}, ``Ecological networks and their
			fragility,'' {\em Nature}, vol.~442, no.~7100, pp.~259--264, 2006.

		\bibitem{ings2009ecological}
			T.~C. Ings, J.~M. Montoya, J.~Bascompte, N.~Bl{\"u}thgen, L.~Brown, C.~F.
			Dormann, F.~Edwards, D.~Figueroa, U.~Jacob, J.~I. Jones, {\em et~al.},
			``Ecological networks--beyond food webs,'' {\em Journal of animal ecology},
			vol.~78, no.~1, pp.~253--269, 2009.

		\bibitem{bascompte2003nested}
			J.~Bascompte, P.~Jordano, C.~J. Meli{\'a}n, and J.~M. Olesen, ``The nested
			assembly of plant--animal mutualistic networks,'' {\em Proceedings of the
			National Academy of Sciences}, vol.~100, no.~16, pp.~9383--9387, 2003.

		\bibitem{corominas2009ontogeny}
			B.~Corominas-Murtra, S.~Valverde, and R.~Sol{\'e}, ``The ontogeny of scale-free
			syntax networks: phase transitions in early language acquisition,'' {\em
			Advances in Complex Systems}, vol.~12, no.~03, pp.~371--392, 2009.

		\bibitem{goni2011semantic}
			J.~Go{\~n}i, G.~Arrondo, J.~Sepulcre, I.~Martincorena, N.~V{\'e}lez~de
			Mendiz{\'a}bal, B.~Corominas-Murtra, B.~Bejarano, S.~Ardanza-Trevijano,
			H.~Peraita, D.~P. Wall, {\em et~al.}, ``The semantic organization of the
			animal category: evidence from semantic verbal fluency and network theory,''
			{\em Cognitive processing}, vol.~12, pp.~183--196, 2011.

		\bibitem{sole2015ambiguity}
			R.~V. Sol{\'e} and L.~F. Seoane, ``Ambiguity in language networks,'' {\em The
			Linguistic Review}, vol.~32, no.~1, pp.~5--35, 2015.

		\bibitem{seoane2018morphospace}
			L.~F. Seoane and R.~Sol{\'e}, ``The morphospace of language networks,'' {\em
			Scientific reports}, vol.~8, no.~1, p.~10465, 2018.

		\bibitem{corominas2018chromatic}
			B.~Corominas-Murtra, M.~S{\`a}nchez~Fibla, S.~Valverde, and R.~Sol{\'e},
			``Chromatic transitions in the emergence of syntax networks,'' {\em Royal
			Society open science}, vol.~5, no.~12, p.~181286, 2018.

		\bibitem{valverde2007topology}
			S.~Valverde, R.~V. Sol{\'e}, M.~A. Bedau, and N.~Packard, ``Topology and
			evolution of technology innovation networks,'' {\em Physical Review E},
			vol.~76, no.~5, p.~056118, 2007.

		\bibitem{valverde2015punctuated}
			S.~Valverde and R.~V. Sol{\'e}, ``Punctuated equilibrium in the large-scale
			evolution of programming languages,'' {\em Journal of The Royal Society
			Interface}, vol.~12, no.~107, p.~20150249, 2015.

		\bibitem{milgram1967small}
			S.~Milgram, ``The small world problem,'' {\em Psychology today}, vol.~2, no.~1,
			pp.~60--67, 1967.

		\bibitem{travers1977experimental}
			J.~Travers and S.~Milgram, ``An experimental study of the small world
			problem,'' in {\em Social networks}, pp.~179--197, Elsevier, 1977.

		\bibitem{neal2014backbone}
			Z.~Neal, ``The backbone of bipartite projections: Inferring relationships from
			co-authorship, co-sponsorship, co-attendance and other co-behaviors,'' {\em
			Social Networks}, vol.~39, pp.~84--97, 2014.

		\bibitem{andris2015rise}
			C.~Andris, D.~Lee, M.~J. Hamilton, M.~Martino, C.~E. Gunning, and J.~A. Selden,
			``The rise of partisanship and super-cooperators in the us house of
			representatives,'' {\em PloS one}, vol.~10, no.~4, p.~e0123507, 2015.

		\bibitem{neal2020sign}
			Z.~P. Neal, ``A sign of the times? weak and strong polarization in the us
			congress, 1973--2016,'' {\em Social Networks}, vol.~60, pp.~103--112, 2020.

		\bibitem{hohmann2023quantifying}
			M.~Hohmann, K.~Devriendt, and M.~Coscia, ``Quantifying ideological polarization
			on a network using generalized euclidean distance,'' {\em Science Advances},
			vol.~9, no.~9, p.~eabq2044, 2023.

		\bibitem{onnela2003dynamics}
			J.-P. Onnela, A.~Chakraborti, K.~Kaski, J.~Kertesz, and A.~Kanto, ``Dynamics of
			market correlations: Taxonomy and portfolio analysis,'' {\em Physical Review
			E}, vol.~68, no.~5, p.~056110, 2003.

		\bibitem{bialek2007rediscovering}
			W.~Bialek and R.~Ranganathan, ``Rediscovering the power of pairwise
			interactions,'' {\em arXiv preprint arXiv:0712.4397}, 2007.

		\bibitem{corominas2013origins}
			B.~Corominas-Murtra, J.~Go{\~n}i, R.~V. Sole, and C.~Rodr{\'\i}guez-Caso, ``On
			the origins of hierarchy in complex networks,'' {\em Proceedings of the
			National Academy of Sciences}, vol.~110, no.~33, pp.~13316--13321, 2013.

		\bibitem{milo2002network}
			R.~Milo, S.~Shen-Orr, S.~Itzkovitz, N.~Kashtan, D.~Chklovskii, and U.~Alon,
			``Network motifs: simple building blocks of complex networks,'' {\em
			Science}, vol.~298, no.~5594, pp.~824--827, 2002.

		\bibitem{newman2006finding}
			M.~E. Newman, ``Finding community structure in networks using the eigenvectors
			of matrices,'' {\em Physical review E}, vol.~74, no.~3, p.~036104, 2006.

		\bibitem{peixoto2021descriptive}
			T.~P. Peixoto, ``Descriptive vs. inferential community detection in networks:
			pitfalls, myths, and half-truths,'' {\em arXiv preprint arXiv:2112.00183},
			2021.

		\bibitem{watts1998collective}
			D.~J. Watts and S.~H. Strogatz, ``Collective dynamics of `small-world'
			networks,'' {\em Nature}, vol.~393, no.~6684, pp.~440--442, 1998.

		\bibitem{manrubia2022report}
			S.~Manrubia~Cuevas, I.~Atienza-Diez, and L.~F. Seoane, ``Report scientific
			collaboration network, 2016-2021 national centre for biotechnology (csic),''
			2022.

		\bibitem{csermely2013structure}
			P.~Csermely, A.~London, L.-Y. Wu, and B.~Uzzi, ``Structure and dynamics of
			core/periphery networks,'' {\em Journal of Complex Networks}, vol.~1, no.~2,
			pp.~93--123, 2013.

		\bibitem{rombach2014core}
			M.~P. Rombach, M.~A. Porter, J.~H. Fowler, and P.~J. Mucha, ``Core-periphery
			structure in networks,'' {\em SIAM Journal on Applied mathematics}, vol.~74,
			no.~1, pp.~167--190, 2014.

		\bibitem{hebert2016multi}
			L.~H{\'e}bert-Dufresne, J.~A. Grochow, and A.~Allard, ``Multi-scale structure
			and topological anomaly detection via a new network statistic: The onion
			decomposition,'' {\em Scientific reports}, vol.~6, no.~1, p.~31708, 2016.

		\bibitem{marcelino2012critical}
			J.~Marcelino and M.~Kaiser, ``Critical paths in a metapopulation model of h1n1:
			Efficiently delaying influenza spreading through flight cancellation,'' {\em
			PLoS currents}, vol.~4, 2012.

		\bibitem{networkResources}
			``Resources for network science by markus keiser's dynamic connectome lab.''
			\url{https://sites.google.com/view/dynamicconnectomelab/resources}.
			\newblock Accessed: 2023-08-28.

		\bibitem{guimera2005worldwide}
			R.~Guimera, S.~Mossa, A.~Turtschi, and L.~N. Amaral, ``The worldwide air
			transportation network: Anomalous centrality, community structure, and
			cities' global roles,'' {\em Proceedings of the National Academy of
			Sciences}, vol.~102, no.~22, pp.~7794--7799, 2005.

		\bibitem{humanConnectome}
			``The pit group connectomes.''
			\url{https://braingraph.org/cms/download-pit-group-connectomes/}.
			\newblock Accessed: 2023-08-28.

		\bibitem{kerepesi2017braingraph}
			C.~Kerepesi, B.~Szalkai, B.~Varga, and V.~Grolmusz, ``The braingraph. org
			database of high resolution structural connectomes and the brain graph
			tools,'' {\em Cognitive Neurodynamics}, vol.~11, pp.~483--486, 2017.

		\bibitem{davidson1995brain}
			R.~J. Davidson and K.~Hugdahl, {\em Brain asymmetry}.
			\newblock Mit Press, 1995.

		\bibitem{seoane2020modeling}
			L.~F. Seoane and R.~Sol{\'e}, ``Modeling brain reorganization after
			hemispherectomy,'' {\em bioRxiv}, pp.~2020--12, 2020.

		\bibitem{carballo2022phase}
			A.~Carballo-Castro and L.~F. Seoane, ``Phase transitions in a simple model of
			focal stroke imitate recovery and suggest neurorehabilitation strategies,''
			{\em bioRxiv}, pp.~2022--12, 2022.

		\bibitem{seoane2020fate}
			L.~F. Seoane, ``Fate of duplicated neural structures,'' {\em Entropy}, vol.~22,
			no.~9, p.~928, 2020.

		\bibitem{seoane2023optimality}
			L.~F. Seoane, ``Optimality pressures toward lateralization of complex brain
			functions,'' {\em Physical Review X}, vol.~13, no.~3, p.~031028, 2023.

		\bibitem{kong2014mapping}
			X.-Z. Kong, S.~R. Mathias, T.~Guadalupe, and E.~Laterality, ``Mapping cortical
			brain asymmetry in 17,141 healthy individuals worldwide via the,'' {\em
			studies}, vol.~40, no.~41, 2014.

		\bibitem{kong2022mapping}
			X.-Z. Kong, M.~C. Postema, T.~Guadalupe, C.~de~Kovel, P.~S. Boedhoe,
			M.~Hoogman, S.~R. Mathias, D.~Van~Rooij, D.~Schijven, D.~C. Glahn, {\em
			et~al.}, ``Mapping brain asymmetry in health and disease through the enigma
			consortium,'' {\em Human brain mapping}, vol.~43, no.~1, pp.~167--181, 2022.

		\bibitem{michaelis2023social}
			A.~C. Michaelis, A.-D. Brunner, M.~Zwiebel, F.~Meier, M.~T. Strauss, I.~Bludau,
			and M.~Mann, ``The social and structural architecture of the yeast protein
			interactome,'' {\em Nature}, vol.~624, no.~7990, pp.~192--200, 2023.

		\bibitem{gallego2017neural}
			J.~A. Gallego, M.~G. Perich, L.~E. Miller, and S.~A. Solla, ``Neural manifolds
			for the control of movement,'' {\em Neuron}, vol.~94, no.~5, pp.~978--984,
			2017.

		\bibitem{houston2022squishy}
			K.~Houston-Edwards, ``Squishy math october 2022, scientificamerican. com
			37topology is becoming an indispensable tool for data analysisand revealing
			doughnuts in the brain,'' 2022.

		\bibitem{gardner2022toroidal}
			R.~J. Gardner, E.~Hermansen, M.~Pachitariu, Y.~Burak, N.~A. Baas, B.~A. Dunn,
			M.-B. Moser, and E.~I. Moser, ``Toroidal topology of population activity in
			grid cells,'' {\em Nature}, vol.~602, no.~7895, pp.~123--128, 2022.

		\bibitem{sebastian2023topological}
			E.~R. Sebastian, J.~P. Quintanilla, A.~S{\'a}nchez-Aguilera, J.~Esparza,
			E.~Cid, and L.~M. de~la Prida, ``Topological analysis of sharp-wave ripple
			waveforms reveals input mechanisms behind feature variations,'' {\em Nature
			neuroscience}, vol.~26, no.~12, pp.~2171--2181, 2023.

		\bibitem{becht2019dimensionality}
			E.~Becht, L.~McInnes, J.~Healy, C.-A. Dutertre, I.~W. Kwok, L.~G. Ng,
			F.~Ginhoux, and E.~W. Newell, ``Dimensionality reduction for visualizing
			single-cell data using umap,'' {\em Nature biotechnology}, vol.~37, no.~1,
			pp.~38--44, 2019.

		\bibitem{fortunato202220}
			S.~Fortunato and M.~E. Newman, ``20 years of network community detection,''
			{\em Nature Physics}, vol.~18, no.~8, pp.~848--850, 2022.

		\bibitem{bianconi2013statistical}
			G.~Bianconi, ``Statistical mechanics of multiplex networks: Entropy and
			overlap,'' {\em Physical Review E}, vol.~87, no.~6, p.~062806, 2013.

		\bibitem{peixoto2022disentangling}
			T.~P. Peixoto, ``Disentangling homophily, community structure, and triadic
			closure in networks,'' {\em Physical Review X}, vol.~12, no.~1, p.~011004,
			2022.

		\bibitem{mucha2010community}
			P.~J. Mucha, T.~Richardson, K.~Macon, M.~A. Porter, and J.-P. Onnela,
			``Community structure in time-dependent, multiscale, and multiplex
			networks,'' {\em science}, vol.~328, no.~5980, pp.~876--878, 2010.

		\bibitem{nicosia2013growing}
			V.~Nicosia, G.~Bianconi, V.~Latora, and M.~Barthelemy, ``Growing multiplex
			networks,'' {\em Physical review letters}, vol.~111, no.~5, p.~058701, 2013.

		\bibitem{moore2012analyzing}
			T.~J. Moore, R.~J. Drost, P.~Basu, R.~Ramanathan, and A.~Swami, ``Analyzing
			collaboration networks using simplicial complexes: A case study,'' in {\em
			2012 Proceedings IEEE INFOCOM Workshops}, pp.~238--243, IEEE, 2012.

		\bibitem{patania2017shape}
			A.~Patania, G.~Petri, and F.~Vaccarino, ``The shape of collaborations,'' {\em
			EPJ Data Science}, vol.~6, pp.~1--16, 2017.

		\bibitem{kramer1991nonlinear}
			M.~A. Kramer, ``Nonlinear principal component analysis using autoassociative
			neural networks,'' {\em AIChE journal}, vol.~37, no.~2, pp.~233--243, 1991.

		\bibitem{mcinnes2018umap}
			L.~McInnes, J.~Healy, and J.~Melville, ``Umap: Uniform manifold approximation
			and projection for dimension reduction,'' {\em arXiv preprint
			arXiv:1802.03426}, 2018.

		\bibitem{van2008visualizing}
			L.~Van~der Maaten and G.~Hinton, ``Visualizing data using t-sne.,'' {\em
			Journal of machine learning research}, vol.~9, no.~11, 2008.

		\bibitem{neal42constructing}
			Z.~P. Neal, ``Constructing legislative networks in r using incidentally and
			backbone,'' {\em Connections}, vol.~42, no.~1, pp.~1--9.

		\bibitem{humanConnectomeProject}
			H.~C. Project", ``{The Human Connectome Project}.''
			\url{https://wiki.humanconnectome.org/}, 2008.
			\newblock Accessed: 2024-08-28.

		\bibitem{stephan2001advanced}
			K.~E. Stephan, L.~Kamper, A.~Bozkurt, G.~A. Burns, M.~P. Young, and
			R.~K{\"o}tter, ``Advanced database methodology for the collation of
			connectivity data on the macaque brain (cocomac),'' {\em Philosophical
			Transactions of the Royal Society of London. Series B: Biological Sciences},
			vol.~356, no.~1412, pp.~1159--1186, 2001.

		\bibitem{kotter2004online}
			R.~K{\"o}tter, ``Online retrieval, processing, and visualization of primate
			connectivity data from the cocomac database,'' {\em Neuroinformatics},
			vol.~2, pp.~127--144, 2004.

		\bibitem{incidentally}
			``R package to build us congress bill-cosponsorship networks.''
			https://cran.r-project.org/web/packages/incidentally/vignettes/congress.html.
			\newblock Accessed: 2024-02-12.

		\bibitem{bonacich1987power}
			P.~Bonacich, ``Power and centrality: A family of measures,'' {\em American
			journal of sociology}, vol.~92, no.~5, pp.~1170--1182, 1987.

		\bibitem{newman2018networks}
			M.~Newman, {\em Networks}.
			\newblock Oxford university press, 2018.

		\bibitem{brandes2001faster}
			U.~Brandes, ``A faster algorithm for betweenness centrality,'' {\em Journal of
			mathematical sociology}, vol.~25, no.~2, pp.~163--177, 2001.

		\bibitem{freeman2002centrality}
			L.~C. Freeman {\em et~al.}, ``Centrality in social networks: Conceptual
			clarification,'' {\em Social network: critical concepts in sociology.
			Londres: Routledge}, vol.~1, pp.~238--263, 2002.

		\bibitem{boldi2014axioms}
			P.~Boldi and S.~Vigna, ``Axioms for centrality,'' {\em Internet Mathematics},
			vol.~10, no.~3-4, pp.~222--262, 2014.

		\bibitem{batagelj2003m}
			V.~Batagelj and M.~Zaversnik, ``An o (m) algorithm for cores decomposition of
			networks,'' {\em arXiv preprint cs/0310049}, 2003.

		\bibitem{allard2019percolation}
			A.~Allard and L.~H{\'e}bert-Dufresne, ``Percolation and the effective structure
			of complex networks,'' {\em Physical Review X}, vol.~9, no.~1, p.~011023,
			2019.

		\bibitem{lazega1995structural}
			E.~Lazega, ``Structural holes: the social structure of competition,'' 1995.

		\bibitem{borgatti1997structural}
			S.~P. Borgatti, ``Structural holes: Unpacking burt's redundancy measures,''
			{\em Connections}, vol.~20, no.~1, pp.~35--38, 1997.

		\bibitem{fan2021characterizing}
			T.~Fan, L.~L{\"u}, D.~Shi, and T.~Zhou, ``Characterizing cycle structure in
			complex networks,'' {\em Communications Physics}, vol.~4, no.~1, p.~272,
			2021.

		\bibitem{lind2005cycles}
			P.~G. Lind, M.~C. Gonzalez, and H.~J. Herrmann, ``Cycles and clustering in
			bipartite networks,'' {\em Physical review E}, vol.~72, no.~5, p.~056127,
			2005.

		\bibitem{zhang2008clustering}
			P.~Zhang, J.~Wang, X.~Li, M.~Li, Z.~Di, and Y.~Fan, ``Clustering coefficient
			and community structure of bipartite networks,'' {\em Physica A: Statistical
			Mechanics and its Applications}, vol.~387, no.~27, pp.~6869--6875, 2008.

		\bibitem{burt2004structural}
			R.~S. Burt, ``Structural holes and good ideas,'' {\em American journal of
			sociology}, vol.~110, no.~2, pp.~349--399, 2004.

		\bibitem{hagberg2008exploring}
			A.~Hagberg, P.~Swart, and D.~S~Chult, ``Exploring network structure, dynamics,
			and function using networkx,'' tech. rep., Los Alamos National Lab.(LANL),
			Los Alamos, NM (United States), 2008.

		\bibitem{johnson2010entropic}
			S.~Johnson, J.~J. Torres, J.~Marro, and M.~A. Munoz, ``Entropic origin of
			disassortativity in complex networks,'' {\em Physical review letters},
			vol.~104, no.~10, p.~108702, 2010.

		\bibitem{pearson1901liii}
			K.~Pearson, ``Liii. on lines and planes of closest fit to systems of points in
			space,'' {\em The London, Edinburgh, and Dublin philosophical magazine and
			journal of science}, vol.~2, no.~11, pp.~559--572, 1901.

		\bibitem{lloyd1982least}
			S.~Lloyd, ``Least squares quantization in pcm,'' {\em IEEE transactions on
			information theory}, vol.~28, no.~2, pp.~129--137, 1982.

		\bibitem{gavish2014optimal}
			M.~Gavish and D.~L. Donoho, ``The optimal hard threshold for singular values is
			$4\ \sqrt{3}$,'' {\em IEEE Transactions on Information Theory}, vol.~60,
			no.~8, pp.~5040--5053, 2014.

		\bibitem{pascual2020functionink}
			A.~Pascual-Garc{\'\i}a and T.~Bell, ``functionink: An efficient method to
			detect functional groups in multidimensional networks reveals the hidden
			structure of ecological communities,'' {\em Methods in Ecology and
			Evolution}, vol.~11, no.~7, pp.~804--817, 2020.

	\end{thebibliography}

\end{document}